\documentclass[twocolumn,preprintnumbers,amsmath,amssymb]{revtex4-1}
\usepackage{graphicx}
\usepackage{ulem}
\usepackage[svgnames]{xcolor}
\usepackage{bm}
\usepackage{multirow}
\usepackage{float}
\usepackage{titlesec}
\usepackage{array}
\usepackage{booktabs}

\begin{document}

\title{Intrinsic anomalous Nernst effect amplified by disorder in a half-metallic semimetal}

\author{Linchao Ding$^{1}$, Jahyun Koo$^{2}$, Liangcai Xu$^{1}$, Xiaokang Li$^{1}$, Xiufang Lu$^{1}$, Lingxiao Zhao$^{1}$, Qi Wang$^{3}$, Qiangwei Yin$^{3}$, Hechang Lei$^{3}$, Binghai Yan$^{2,*}$ Zengwei Zhu$^{1,*}$ and Kamran Behnia$^{4, 5, *}$}

\affiliation{(1) Wuhan National High Magnetic Field Center and School of Physics, Huazhong University of Science and Technology,  Wuhan,  430074, China\\
(2) Department of Condensed Matter Physics, Weizmann Institute of Science, 7610001 Rehovot, Israel\\
(3)Department of Physics and Beijing Key Laboratory of Opto-electronic Functional Materials \& Micro-nano Devices, Renmin University of China, Beijing, 100872, China\\
(4)Laboratoire de Physique et Etude des Mat\'{e}riaux (CNRS/UPMC),
Ecole Sup\'{e}rieure de Physique et de Chimie Industrielles, 10 Rue Vauquelin, 75005 Paris, France\\
(5) II. Physikalisches Institut, Universit\"{a}t zu K\"{o}ln, 50937 K\"{o}ln, Germany
}

\date{\today}

\begin{abstract}
Intrinsic anomalous Nernst effect (ANE), like its Hall counterpart, is generated by Berry curvature of electrons in solids. Little is known about its response to disorder. In contrast, the link between the amplitude of the ordinary Nernst coefficient (ONE) and the mean-free-path is extensively documented. Here, by studying Co$_3$Sn$_2$S$_2$, a topological half-metallic semimetal hosting sizable and recognizable ordinary and anomalous Nernst responses, we demonstrate an anti-correlation between the amplitude of ANE and carrier mobility. We argue that the observation, paradoxically, establishes the intrinsic origin of the ANE in this system. We conclude that various intrinsic off-diagonal coefficients are set by the way the Berry curvature is averaged on a grid involving the mean-free-path, the Fermi wavelength and the de Broglie thermal length.

\end{abstract}
\maketitle

\section{Introduction}
Electrons in some solids do not flow along the applied electric field or thermal gradient, even in zero magnetic field. The phenomena, known as anomalous Hall, anomalous Nernst and anomalous thermal Hall effects (AHE, ANE and ATHE), may be caused either by the geometric (Berry) curvature of Bloch functions or by the skew scattering of electrons off magnetic impurities. Distinguishing between these intrinsic and extrinsic origins (see \cite{Nagaosa2010,Xiao2010} for reviews) has motivated numerous investigations during the past two decades.

Here we focus on the anomalous Nernst effect (ANE). The Nernst effect, $S_{xy}$, is a transverse electric field generated by a longitudinal temperature gradient (in absence of charge flow). It is directly accessible to the experimentalist and has to be distinguished from the off-diagonal component of the thermoelectric tensor, ($\alpha_{xy}$). The latter is the transverse temperature gradient generated by a longitudinal charge flow (in absence of electric field), which is often the immediate result of theoretical calculations. The two coefficients are intimately connected to each other \cite{Behnia2016}.

A pioneer theoretical study   of the \textit{intrinsic} anomalous Nernst effect \cite{Xiao2006} argued that it depends on the magnitude of the magnetization in the host solid. The expected correlation between the two properties became the mantra of many of the numerous experimental studies of the ANE \cite{Miyasato2007,Sakai2018,Lee2004,Miyasato2007,PuYong2008, Ikhlas2017,Li2017,Xu2018,Yuke2018,Wuttke2019,Guin2019} in topological solids (for a review of such systems see  \cite{Armitage2018review,Yan2017review}). A prominent question, yet to be addressed, is the role of disorder in setting the amplitude of the ANE. Does this transport property depend on the mean-free-path? If yes, since magnetization is a thermodynamic property, how come? The ordinary Nernst effect (ONE) is expected to scale with mobility \cite{Behnia2016,Behnia2009}. Experimental studies have not only confirmed this scaling across different systems, but also found that the magnitude of ONE in a given solid increases as it becomes cleaner (bismuth and URu$_2$Si$_2$ are two prominent case studies) \cite{Behnia2016}. To the best of our knowledge, this has not been the case of ANE.

In this paper, we show that this opportunity is offered for the first time by the newly-discovered magnetic Weyl semimetal Co$_3$Sn$_2$S$_2$. This is a solid with a Shandite structure, which becomes a ferromagnet below 180 K \cite{Vaqueiro2009,Holder2009, Schnelle2013}. Theoretical calculations suggest that this semimetallic half-metal \cite{Coey2004} with negative flat-band magnetism \cite{Yin2019}, hosts Weyl nodes 60 meV off the Fermi level \cite{LiuEnke2018}. Previous experiments have found a large AHE \cite{LiuEnke2018,Lei2018}, a large ANE \cite{Yuke2018,Guin2019} and surface-termination-dependent Fermi arcs\cite{Morali2019}.

We report first on a significant improvement in the quality of Co$_3$Sn$_2$S$_2$ single crystals obtained by chemical vapor transport (CVT) method. The improved quality shows itself in higher mobility compared to what was reported in previous studies \cite{Schnelle2013,LiuEnke2018,Lei2018,Yuke2018,Guin2019}. This allowed us to perform a systematic study of five different crystals with different impurity concentrations. We find that, as expected \cite{Behnia2009,Behnia2016}, the amplitude of the ordinary Nernst response, $S_{xy}\rm ^O$ is proportional to mobility, $\mu$; On the other hand, the amplitude of the anomalous Nernst effect, $S{_{xy}\rm ^A}$ is proportional to the inverse of $\mu$.  We will argue then the amplification of $S{_{xy}\rm ^A}$ by disorder reflects the fact that the anomalous transverse thermoelectricity, $\alpha{_{xy}\rm ^A}$ depends on the Berry curvature (averaged over a reciprocal distance set by the thermal de Broglie wavelength), but not on the mean-free-path. This is to be contrasted with the  semi-classical $\alpha_{xy}$, which scales with the square of the mean-free-path \cite{Behnia2016}. We show that according to both theory and experiment, the Fermi surface of Co$_3$Sn$_2$S$_2$ is complex. Finally, we determine the magnitude of the anomalous transverse thermoelectric conductivity,  $\alpha{_{xy}\rm ^A}$ in all five samples and find a qualitative agreement between theory and experiment with no need for invoking  uncontrolled and unidentified extrinsic dopants invoked previously \cite{Guin2019}. It is remarkable that the scaling between average mobility and the anomalous Nernst coefficient remains valid in spite of the  multiplicity of Fermi surface pockets.

\begin{figure}
\includegraphics[width=9cm]{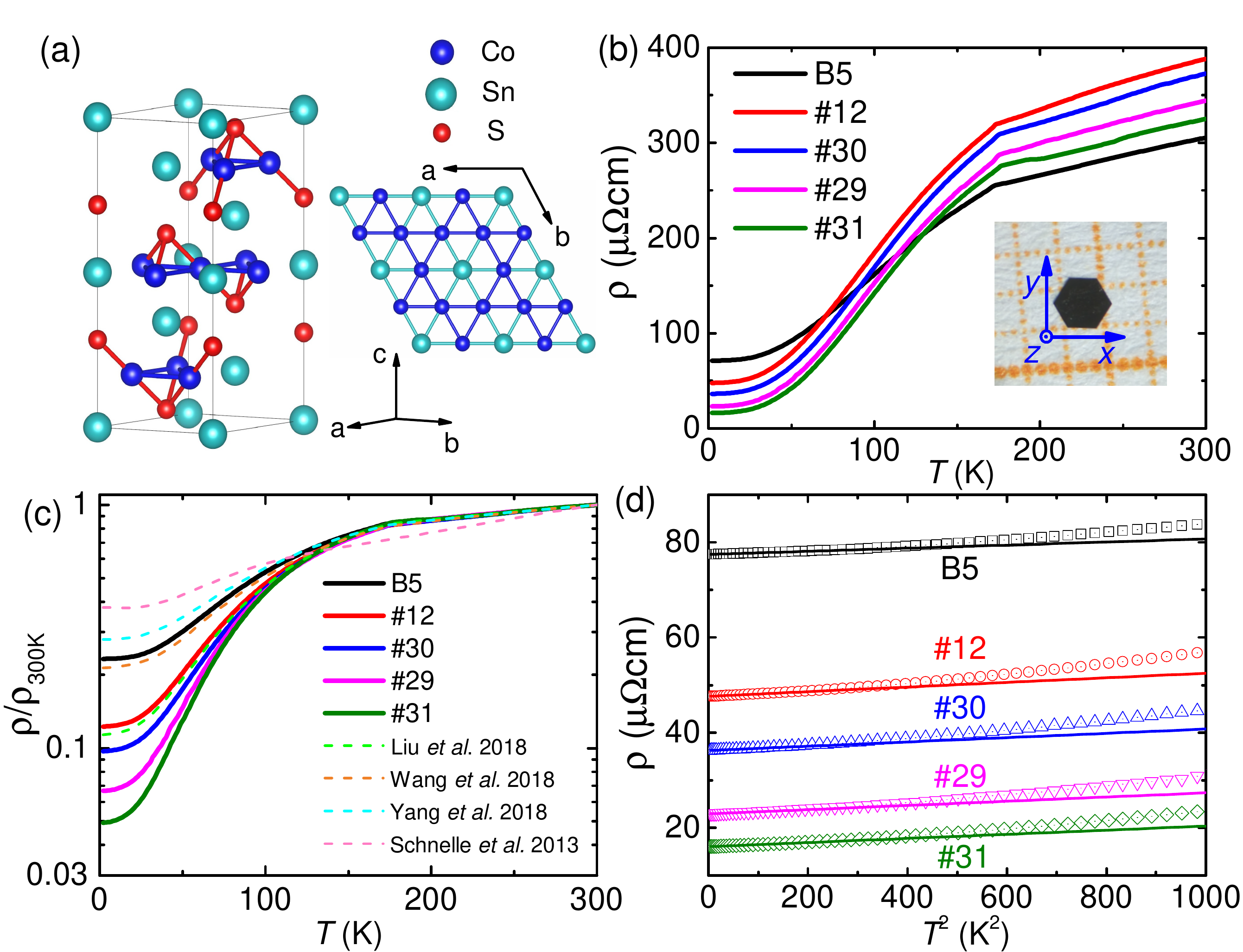}
\caption{ \textbf{Crystal structure, resistivity and sample-dependent disorder-}(a) Crystal structure of Co$_3$Sn$_2$S$_2$. (b) The temperature dependence of resistivity at zero-field in five samples. The residual resistivity ratio (RRR) ranges from 4 to 21.  The inset photograph shows an as-grown hexagonal-shape sample with a typical dimension of 1.3$\times$1$\times$0.02 mm$^3$. The orientations of x, y and z are defined as [2\={1}\={1}0], [01\={1}0] and [0001] respectively. (c) Comparison of resistivity, normalized to its room-temperature value between the samples used in this study and four previous reports. The previous RRR is comparable with the two dirtiest sample in this study. (d) $\rho$ as the function $T^2$ in our samples. The prefactor of T-square resistivity is similar in all fours samples grown from chemical vapor transport method but the difference is residual resistivity leads to an upward shift. }
\label{basicinfo}
\end{figure}

\section{Samples with different mobilities}

Co$_3$Sn$_2$S$_2$  crystallizes in the shandite structure, which consists of ABC stacking of Kagome sheets (See Fig. \ref{basicinfo}(a)). A picture of a hexagonal sample with typical dimension of 1.3$\times$1$\times$0.02 mm$^3$ is shown in the inset of Fig. \ref{basicinfo}(b). The in-plane resistivity, $\rho_{xx}$, of the five samples (among a dozen measured \cite{SM}) is shown in Fig. \ref{basicinfo}(b). A  kink  at 177 K is visible in $\rho_{xx}(T)$. It indicates the Curie temperature and the opening of a gap in the spectrum of minority spins leading to the half metallicity \cite{Vaqueiro2009,Holder2009}.  The residual resistivity  ratio (RRR = $\rho(300 \rm{K})/\rho(2 \rm{K})$) varies between 4 in the dirtiest sample (B5) and 21 in the cleanest (\#31). Fig. \ref{basicinfo}(c), compares the normalized resistivity (i.e. $\rho(T)/\rho(300 \rm{K})$ of the samples with the data reported in four previous studies \cite{Schnelle2013, LiuEnke2018,Lei2018,Yuke2018}. One can see that three of our chemical-vapor-transport (CVT) samples display the highest RRR. In Fig. \ref{basicinfo}(d), the low-temperature resistivity is plotted \textit{vs.} $T^2$. One can see that disorder shifts the curves rigidly upwards and the five-fold change in residual resistivity barely affects inelastic scattering. Indeed, the prefactor of the T-square resistivity \cite{Lin2015} is the same (4.5 n$\Omega$cmK$^{-2}$) (see the supplement \cite{SM} for details).

\begin{figure}
\includegraphics[width=8cm]{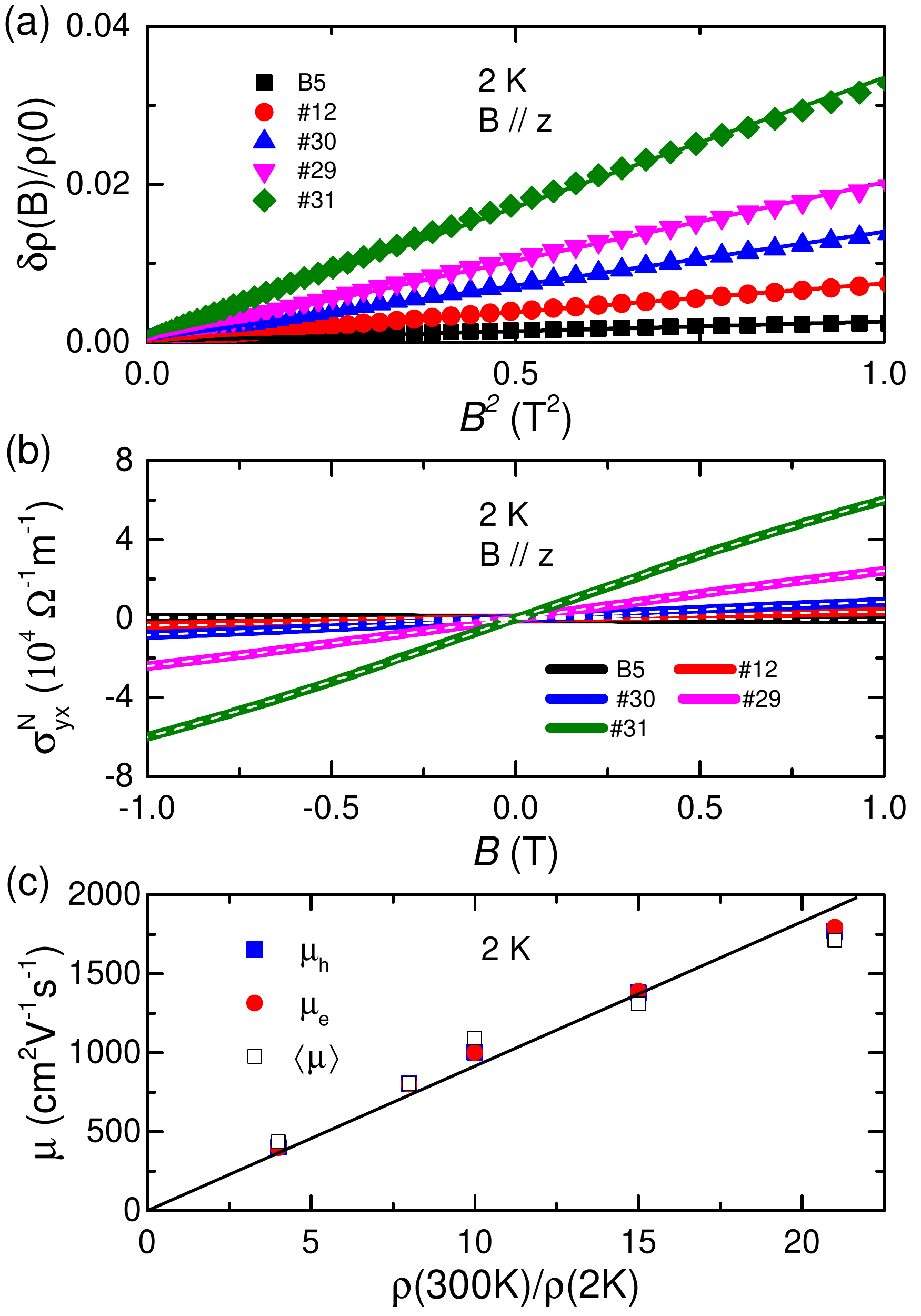}
\caption{ \textbf{Quantifying mobility-} (a) Low-field magnetoresistance in five different samples as a function of $B^2$(T$^2$), where $\delta\rho(B)=\rho(B)-\rho(0)$. As expected, magnetoresistance is quadratic in field and is larger in cleaner samples. (b) Low-field ordinary Hall conductivity. Dashed lines show fits to a two-band model. (c) Mobility of electrons and holes (extracted from Hall conductivity), the average mobility (extracted from magnetoresistance) as a function of RRR of each sample. They scatter around solid line with zero intercept. }
\label{mobility}
\end{figure}

Average mobility and carrier concentration in each sample can be quantified using magnetoresistance (Fig.\ref{mobility}(a)) and the ordinary Hall conductivity (Fig.\ref{mobility}(b))  (See \cite{SM} for details). The average mobility extracted from magnetoresistance is in good agreement with the mobility of electrons and holes extracted from the fit to the ordinary Hall conductivity. The carrier density for electrons and holes was found to be $n_e(\cong n_h) =8.7\pm0.3 \times 10^{19}cm^{-3}$. As one can see in (Fig.\ref{mobility}(c)), the mobilities scale with the magnitude of RRR, providing another check of our assumption that what distinguishes these crystals is the disorder-limited mobility of carriers.



\section{Evolution of anomalous and ordinary responses with disorder}
\begin{figure}
\includegraphics[width=9cm]{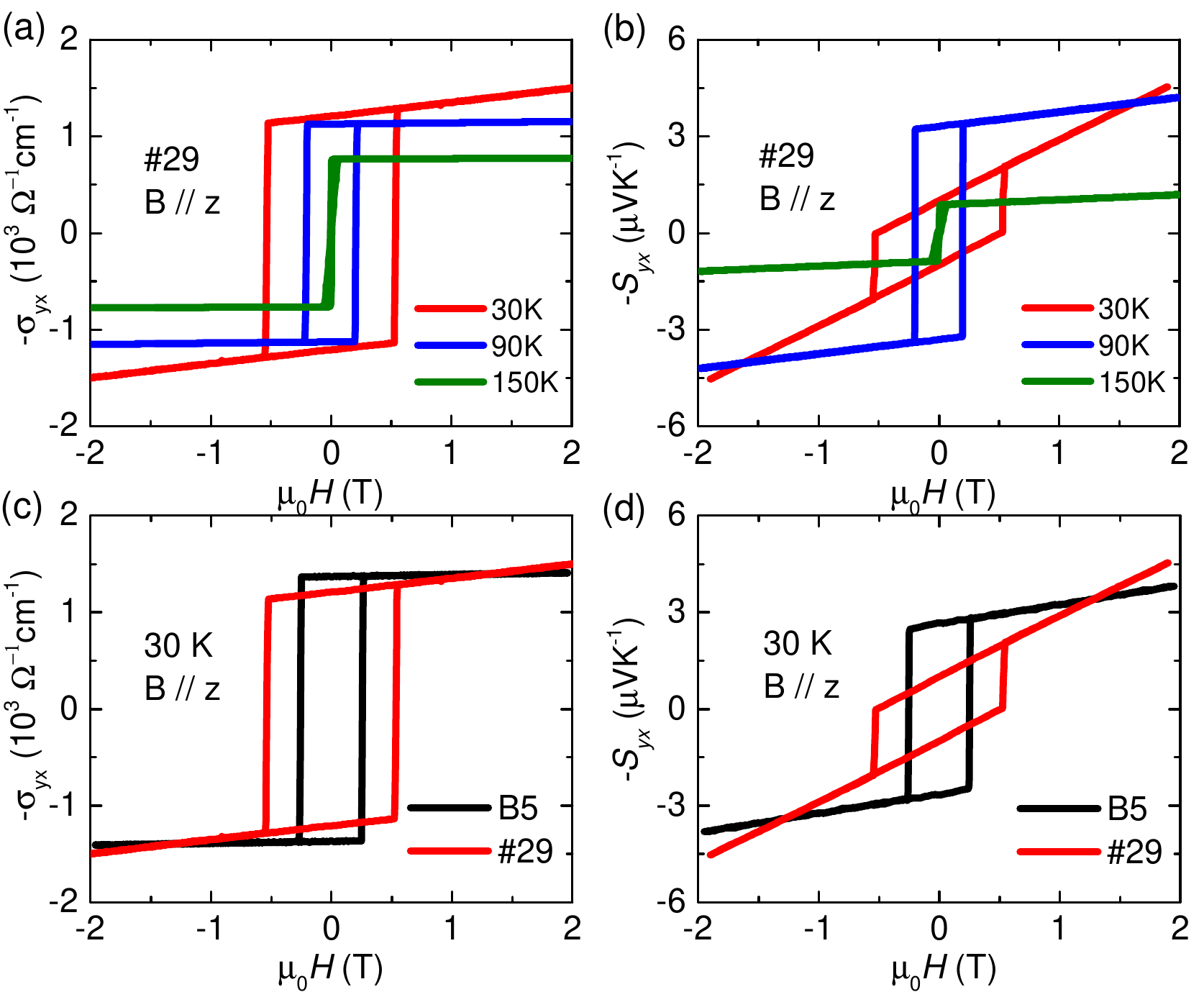}
\caption{ \textbf{Temperature-dependent AHE and ANE and their response to disorder-} Top: Temperature dependence: (a) The Hall conductivity as a function of  magnetic field at three different temperatures; (b) The  Nernst coefficient as a function of magnetic field in at three different temperatures. Note the emergence of a hysteretic loop and anomalous responses in the ferromagnetic state. Bottom:  The effect of disorder: (c) The Hall conductivity compared in two samples (the clean \#29 and the dirty B5) at 30 K.  (d). The Nernst coefficient in the same two samples at the same temperature. The jump, which represents the anomalous component is larger in the dirtier sample, but the slope which represents the ordinary component is larger in the clean sample. Note the wider hysteretic loop in the cleaner sample in both Hall and Nernst data.}
\label{disorder}
\end{figure}

The field dependence of Hall conductivity $\sigma_{yx}$ and the  Nernst signal $S_{yx}$ in sample \#29 at different temperatures are shown in the top panels (a, b) of Fig. \ref{disorder}. As reported  previously \cite{LiuEnke2018,Lei2018,Yuke2018}, there is  a large hysteretic jump in both coefficients in the ferromagnetic state of  Co$_3$Sn$_2$S$_2$.  The  magnitude of the anomalous Hall conductivity is only slightly larger than what observed in BCC-iron \cite{Li2017}. On the other hand, the anomalous Hall angle becomes as large as 20\%, which as previously noticed \cite{LiuEnke2018} is a record value. This  arises because of the low longitudinal conductivity in this dilute metal.

In the bottom panels (c) and (d) of the same Fig. \ref{disorder}, one can see how  disorder affects the two transport coefficients. In the Hall channel, the most visible difference between a clean and a dirty sample is the width of the hysteresis loop. There is a difference in the slope (which represents the ordinary component of the Hall conductivity set by mobility), but a very slight variation in the height of the jump (which quantifies the size of the anomalous component). In the case of the Nernst response, the difference is much more striking  (Fig. \ref{disorder}d). The cleaner sample presents a much smaller jump pointing to the amplification of the anomalous component with disorder.

\begin{figure*}
\includegraphics[width=17cm]{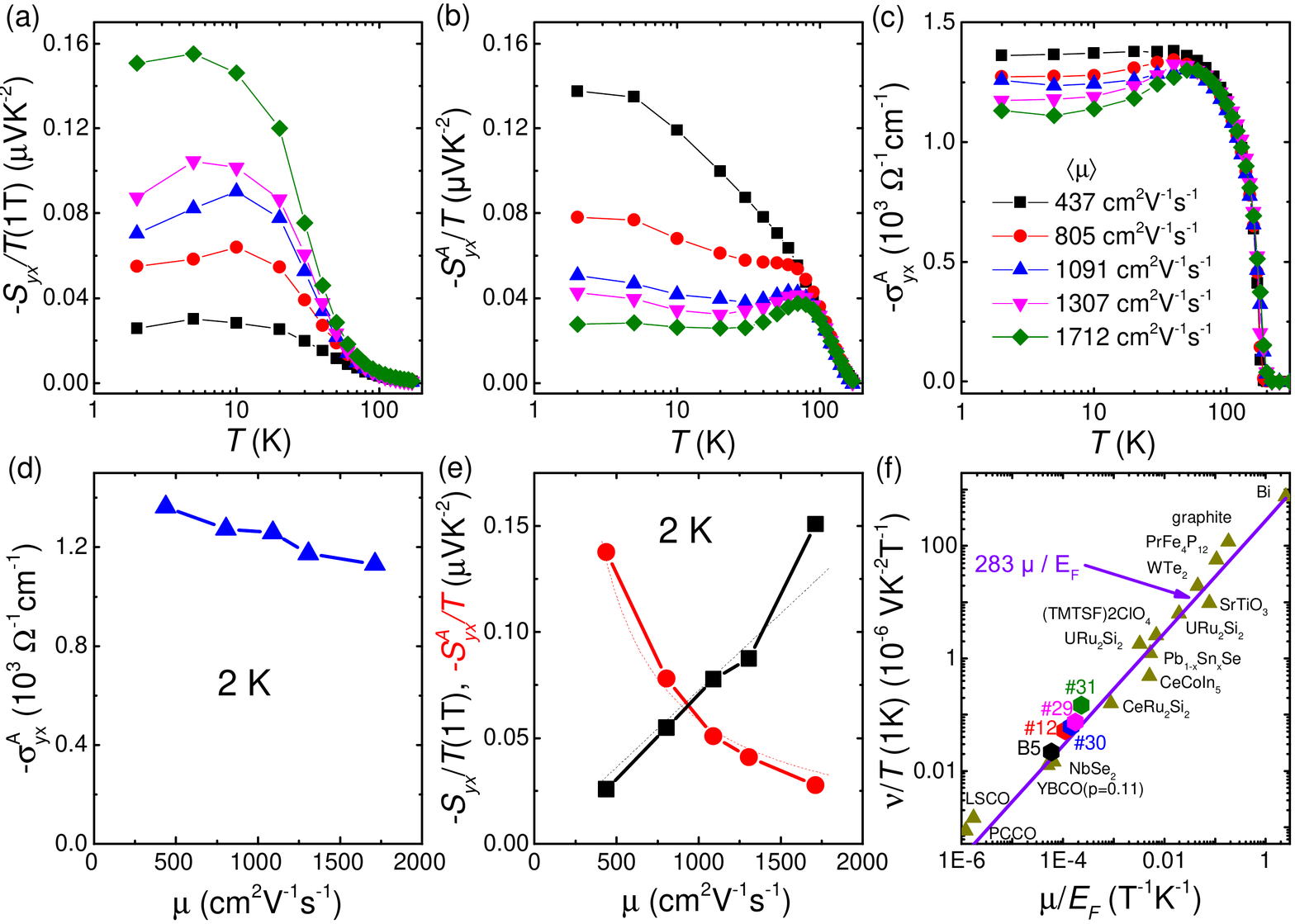}
\caption{ \textbf{Opposite evolution of ordinary and anomalous Nernst effects with disorder} a, b, c) Temperature dependence of the anomalous Hall conductivity $\sigma{_{yx}\rm ^A}$ (a); the anomalous Nernst effect $S{_{yx}\rm ^A}$ (b) and the  ordinary Nernst  effect  $S{_{yx}\rm ^O/T}$ (c) in five different samples with different mobilities as defined in panel a. (d) Low-temperature AHC modestly decreases with incraesing mobility. (e) Variation of  $S{_{yx}\rm ^O}$ and $S{_{yx}\rm ^A}$ with mobility. the first is proportional to mobility (dashed black line), but the second proportional to the inverse of mobility (dashed red line). (f) The ordinary Nernst coefficient $\nu=S{_{xy}\rm ^O}/B$ at 2 K compared with other solids. The solid line represents the amplitude expected in the semi-classical picture ($\nu/T=\frac{\pi^2}{3}\frac{k_B}{e}\frac{\mu}{E_F}$) \cite{Behnia2016}.}
\label{Fig:ONEANE}
\end{figure*}

We carried out an extensive set of measurements on the five samples with different mobilities in order to see how the transport coefficients evolve with temperature. A slightly attenuated disorder dependence is visible in the case of $\sigma{_{yx}\rm ^A}$ in Fig. \ref{Fig:ONEANE}c. The five-fold increase in mobility leads to a 10 percent decrease in the magnitude of low-temperature $\sigma{_{yx}\rm ^A}$ (Fig. \ref{Fig:ONEANE}d). Fig. \ref{Fig:ONEANE}b shows the temperature dependence of $S^A_{yx}$ extracted from the jump. Fig. \ref{Fig:ONEANE}c shows the same for ONE $S{_{yx}\rm ^O}/T$ in different samples, extracted from the $S_{yx}$ following its anomalous jump.  Comparing the two plots, one can easily see the opposing tendencies.  In the low-temperature limit, ONE becomes larger in the clean samples while the opposite is true for ANE.   Plotting the low-temperature $S{_{yx}\rm ^A}$ and  $S{_{yx}\rm ^O}$ as a function of mobility (panel e) reveals a striking contrast. One scales with $\mu$, while the other scales with $\mu^{-1}$. As one can see in panel (f), which compares the magnitude of ONE in the system under study with other solids, the magnitude of ONE is consistent with what is expected in a semiclassical picture given the mobility and the Fermi energy. Let us now turn our attention to the magnitude of ANE.


\section{What sets the amplitude of the intrinsic anomalous Nernst effect?}
At first sight, one may think that the amplification of the anomalous Nernst response by disorder points to an extrinsic origin. We now proceed to show that the opposite is true and this is a signature of intrinsic Berry-curvature based origin. We begin by recalling the expression for Hall conductivity for a two-dimensional circular isotropic Fermi surface:

\begin{equation}\label{1}
\sigma^{2D}_{xy} = \frac{ne\mu}{1+\mu^2B^2} \mu B= \frac{e^2}{h}\frac{\ell^2}{\ell_B^2} [\frac{1}{1+\ell^2k_F^{-2}\ell_B^{-4}}]
\end{equation}

Here, $\ell$ is the mean-free-path, $k_F$ ($\lambda_F$) is the Fermi wave-vector (wave-length) and $\ell_B=(\hbar/eB)^{1/2}$ is the magnetic length. We neglect numerical factors and focus on physical parameters. Two extreme limits, the low field ($\mu B \ll 1$) and the high field ($\mu B \gg 1$) are to be distinguished:
\begin{equation}\label{2}
\sigma^{LF}_{xy} \approx \frac{e^2}{h} \frac{\ell^{2}}{\ell_B^{2}}
\end{equation}
\begin{equation}\label{2}
\sigma^{HF}_{xy} \approx \frac{e^2}{h}\frac{\ell_B^{2}} {\lambda_F^{2}}
\end{equation}

In the high-field limit  quantum effects dominate. The latter expression provides the basis for the often-used expression for the anomalous Hall conductivity:

\begin{equation}\label{4}
\sigma^{A}_{xy}=-\frac{e^{2}}{\hbar}\int_{BZ}\frac{d^{D}k}{(2\pi)^{D}}f(k)\Omega_B(k)\approx -\frac{e^2}{\hbar} <\frac{\Omega_B}{\lambda_F^2}>
\end{equation}

For simplicity, we focus on the two-dimensional case ($D=2$) and assume that $<\frac{\Omega_B}{\lambda_F^2}>$ represents the summation of the Berry curvature, $\Omega_B(k)$, over the relevant cross section of the Fermi surface. $\Omega_B$ has replaced the square of the magnetic length, $\ell_B^2$. Berry curvature is indeed a fictitious magnetic field.

We turn our attention to the off-diagonal thermoelectric conductivity, $\alpha_{xy}$. According to the Mott's relation, it quantifies the change in $\sigma_{xy}$ caused by an infinitesimal shift in the chemical potential \cite{Behnia2015}.Thus  if the mean-free-path were constant, there would be no low-field $\alpha_{xy}$. However, in any realistic metal, scattering is energy-dependent and therefore (See the supplement \cite{SM}) :

\begin{equation}\label{5}
\alpha^{LF}_{xy} \approx \frac{ek_B}{h} \frac{\ell^{2}}{\ell_B^{2}}\frac{\lambda_F^{2}}{\Lambda^{2}}
\end{equation}


Here $\Lambda=\sqrt{\frac{h^2}{2 \pi m^*k_BT}}$ is the de Broglie thermal wave-vector. The experimentally-measured Nernst signal, S$_{yx}=\frac{E_y}{\nabla_xT}$  is simply the ratio of $\alpha_{xy}$ to longitudinal resistivity (in the small-Hall-angle limit). Thus:
\begin{equation}\label{6}
S^{LF}_{xy} \approx \frac{k_B}{e} \frac{\ell}{\ell_B^{2}}\frac{\lambda_F^{3}}{\Lambda^{2}}
\end{equation}
Thus, the ordinary Nernst signal is proportional to the  mean-free-path. Note that this equation is identical to the statement that the Nernst signal at low temperatures is set by the ratio  of mobility to the Fermi energy \cite{Behnia2016}, which is the case in numerous metals  (Fig. 4f). Let us now consider the high-field regime. In this limit, it is the energy dependence of $\lambda_F$ (and not $\ell$), which matters. Using Eq. 3 and the Mott relation one finds (See \cite{SM}):

\begin{equation}\label{7}
\alpha^{HF}_{xy} \approx \frac{ek_B}{h} \frac{\ell_B^{2}}{\Lambda^2}
\end{equation}

In analogy with $\sigma{_{xy}\rm ^A}$,  this expression can be used to quantify the magnitude of $\alpha{_{xy}\rm ^A}$ caused by the Berry curvature fictitious field:

\begin{equation}\label{8}
\alpha^{A}_{xy} \approx \frac{ek_B}{h} <\frac{\Omega_B}{\Lambda^2}>
\end{equation}

We see that $\alpha{_{xy}\rm ^A}$ is also set by the summation of the Berry curvature over a cross-section of the Fermi surface, with the thermal wavelength replacing the Fermi wavelength\cite {Behnia2015}. The anomalous Nernst effect would be:
\begin{equation}\label{10}
S^{A}_{xy} \approx \frac{k_B}{e} \frac{1}{k_F \ell}<\frac{\Omega_B}{\Lambda^2}>
\end{equation}
This equation implies a $S{_{xy}\rm ^A}$ inversely proportional to the mean-free-path. Thus, both the correlation of the ONE and the anti-correlation of the ANE  with the carrier  mobility can be explained from Berry-curvature point of view.

However, this is a single-band approach and, as we are going to see in the next section, the solid under study has multiple Fermi surface sheets.

\section{Multiplicity of the Fermi surface pockets in Co$_3$Sn$_2$S$_2$}

\begin{figure*}
\includegraphics[width=16cm]{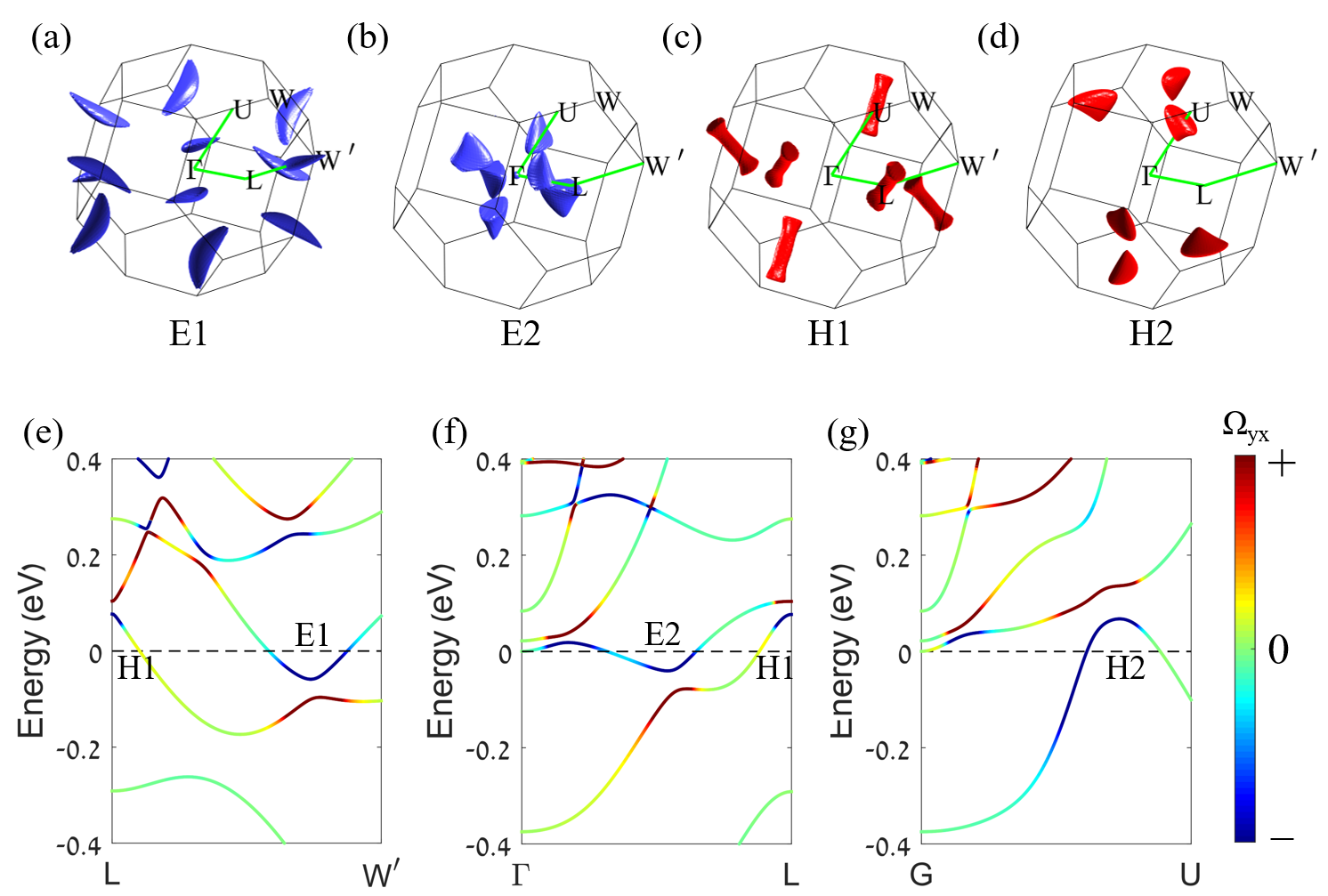}
\caption{\textbf{Theoretical bands and Fermi surfaces.}  The electron and the  hole pockets of Co$_3$Sn$_2$S$_2$. There are two types of electron pockets (a) banana shape pockets called E1 and (b) triangular shape pockets called E2, and two type of hole pockets (c) cylindrical shape pockets called H1 and (d) triangular shape pockets called H2. The lower panels show energy dispersion along three orientations and the location of the pockets follow green lines in upper panels. Colors represent the Berry curvature.}
\label{fig:Fermisurface}
\end{figure*}

According to our theoretical calculations, the Fermi surface of this compensated semi-metal consists of numerous pockets. For both electrons and holes, there are two types of pockets. In the case of electrons there are six banana-shape pockets between L and W¡¯ (called E1), and
six triangle-lik along  $\Gamma$ and L (called E2). In the case of holes, there are three cylinder-like pockets at L (called H1), and six triangle-like pockets between $\Gamma$  and U (called H2) (See Fig. \ref{fig:Fermisurface}). Integrating the size of Fermi pockets leads to the carrier density:
$n_e= n_h=1.7\times 10^{20}$ cm$^{-3}$ for the perfectly stoichiometric solid (that is in absence of uncontrolled doping)\cite{SM}. This  is twice larger than the one extracted experimentally from experimental Hall conductivity and residual resistivity . At the charge neutrality, the distance between the chemical potential and the band edges for electron/hole bands is in the range of 40 to 76 meV.
The calculated cyclotron masses for these pockets are between 0.3 and 1.3 times the bare electron mass (see  table\ref{Table:FS}).

\begin{table}[!hbp]
\begin{tabular}{r|cccc|cc}
\hline
  & \multicolumn{4}{c|}{Theory} & \multicolumn{2}{c}{Experiment}\\
\hline
  &E1& E2& H1 &H2 & Pocket1 & Pocket2 \\
 \hline
F(B//\textit{z}) & 216 & 312 & 114/160 & 402 & 148 & 238 \\
m$^*$(B//\textit{z})& 0.62 & 1.08 & 0.29/0.76 &0.98 &  0.53 & 0.46 \\
F(B//\textit{xy})& 296 & 490 & 701 & 490 & 377 & 586 \\
m$^*$(B//\textit{xy})& 0.86 & 1.21 &1.20 & 1.20 & 0.68 & 1.03 \\
E$_F$(B//z)   & 58   & 40   &77   & 68 &  64 & 66 \\
E$_F$(B//\textit{xy}) &&&&&  33 & 60 \\
\hline
\end{tabular}
\caption{Theoretical and experimental frequencies (F) in unit of telsa, effective masses ($m^*$) in unit of free electron mass $m_e$ for two orientations of magnetic field. The theoretical E$_F$(B//z) in unit of meV is taken from the band edge of corresponding band in Fig.~\ref{fig:Fermisurface}. We note that the calculated H1 pocket has two extreme cross-sections and thus shows two cyclotron frequencies. The cyclotron orbits of each pocket can be found in the supplementary information(Fig. \ref{Fig: ExOrbit}).}
\label{Table:FS}
\end{table}

\begin{figure}
\includegraphics[width=9cm]{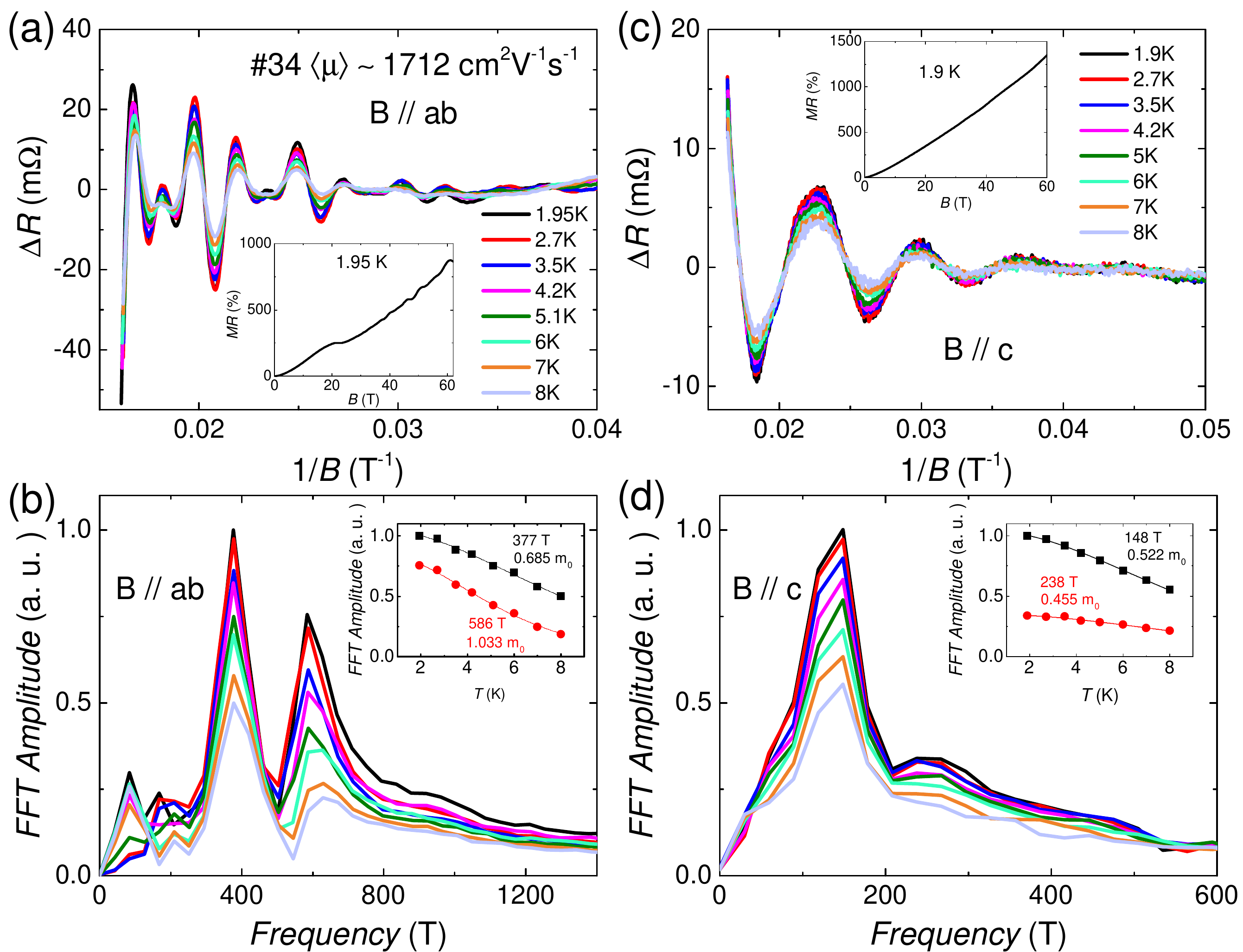}
\caption{ \textbf{Shubnikov-de Haas (SdH) data for $B//ab$ and $B//c$-}   Oscillatory component of magnetoresitance after subtracting a monotonous background for $B//ab$ (a) and for $B//c$. (c).  The FFT  shows two peaks for both configurations (b) and (d). The inset shows the way effective masses of different frequencies were extracted from using a Lifshitz-Kosevich fit to the temperature dependence of the amplitude of the FFT peaks.}
\label{SdH}
\end{figure}

Our experimental study of magnetoresistance up to 60 T for two field orientations ($B//ab$ and $B//c$) led to the observation of quantum oscillations (See Fig. \ref{SdH}). The oscillatory part of the magnetoresistance is visible after subtracting the background (Fig. \ref{SdH}a,b). Performing a Fast Fourier Transform, one can resolve two principal frequencies (Fig. \ref{SdH}c,d). The temperature dependence of the oscillations allow to extract the corresponding masses of each frequency. The frequencies, effective masses and the Fermi energies  are summarized in table \ref{Table:FS}.

 We obtain the theoretical SdH frequencies by the Onsager relation $F=(\Phi_0/2\pi^2)A$, where $A$ is the extreme cross-section of corresponding Fermi pocket, $\Phi_0=h/2e$ is the magnetic flux quantum and $h$ is the Planck constant. One can see that the experimental and theoretical frequencies roughly agree.

At this stage and in the absence of a detailed angle-dependent study of the SdH frequencies, we cannot reach to a definite identification of the electron and hole pockets postulated by theory. Thus, the definite Fermiology of this solid remains unestablished and a task for future studies. Nevertheless, what previous studies pointed out \cite{Schnelle2013,LiuEnke2018} is confirmed by our study. There are multiple sheets to the Fermi surface of this compensated semi-metal.

\section{Anomalous off-diagonal thermoelectric conductivity,  $\alpha{_{xy}\rm ^A}$: theory and experiment}

Previous studies of the anomalous Nernst effect in this system \cite{Guin2019,Yuke2018} reported on the magnitude of the anomalous off-diagonal thermoelectric conductivity, $\alpha{_{xy}\rm ^A}(T)$. As we will see below,  our data and analysis lead us to a different picture from those.

\begin{figure}
\includegraphics[width=9cm]{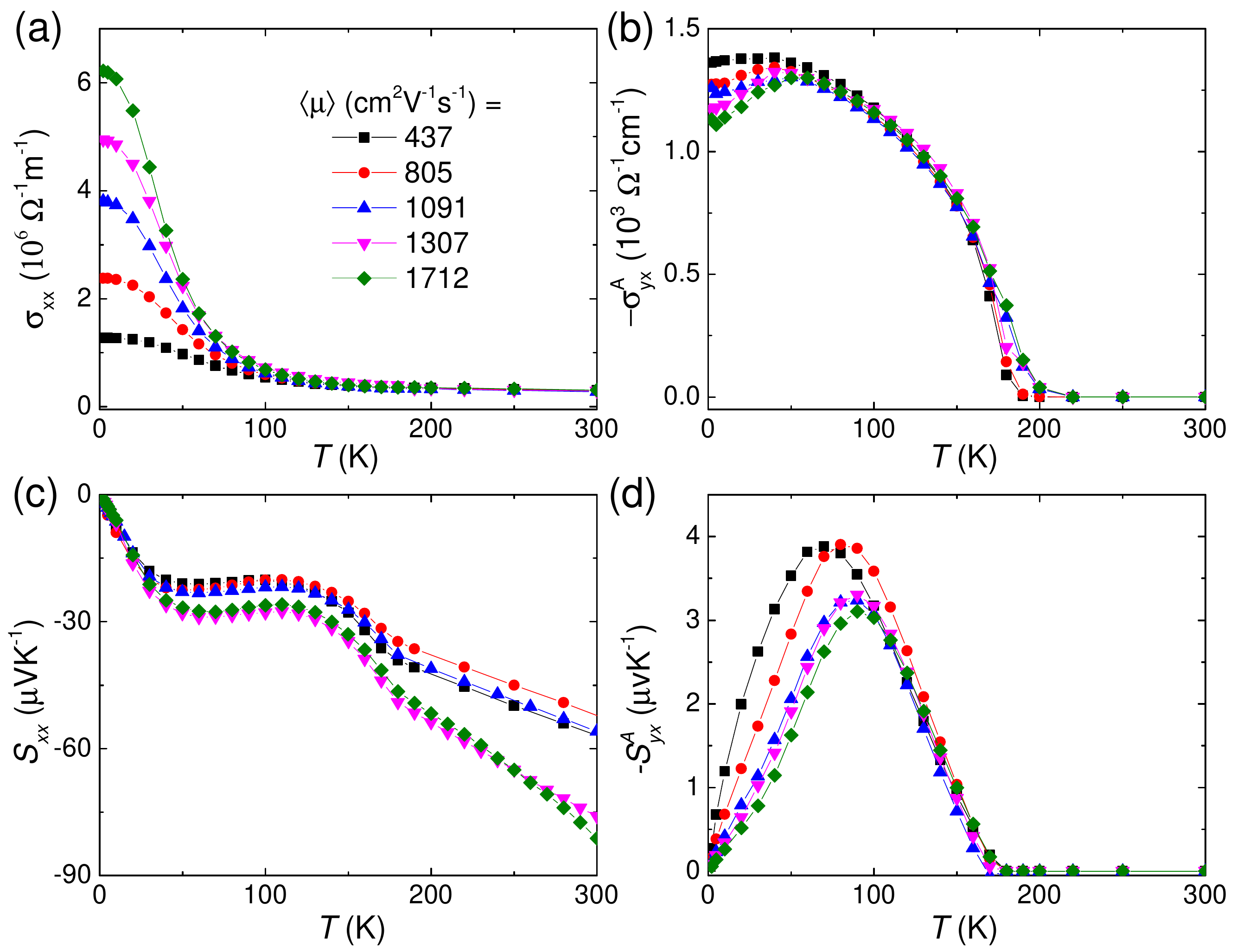}
\caption{\textbf{Four different transport coefficients and disorder-} Temperature dependence of (a) longitudinal  conductivity; (b) the Seebeck coefficient; (c) anomalous Hall conductivity; and (d) The anomalous Nernst coefficient in the five samples. Note the strong variation of longitudinal conductivity and the anomalous Nernst coefficient with disorder, the mild effect on anomalous Hall conductivity and the constancy of the Seebeck coefficient in the low -temperature limit.}
\label{4coeff.}
\end{figure}

The anomalous off-diagonal thermoelectric conductivity is linked to the anomalous Nernst coefficient, $S{_{xy}\rm ^A}$ through three other coefficients, which are the  anomalous Hall conductivity  $\sigma{_{xy}\rm ^A}$, the (ordinary) longitudinal conductivity, $\sigma_{xx}$, and the (ordinary) Seebeck coefficient,$S_{xx}$:

\begin{equation}\label{10}
\alpha^{A}_{xy}=S_{xx}\sigma^{A}_{xy} + S^{A}_{xy} \sigma_{xx}
\end{equation}
\begin{figure*}
\includegraphics[width=18cm]{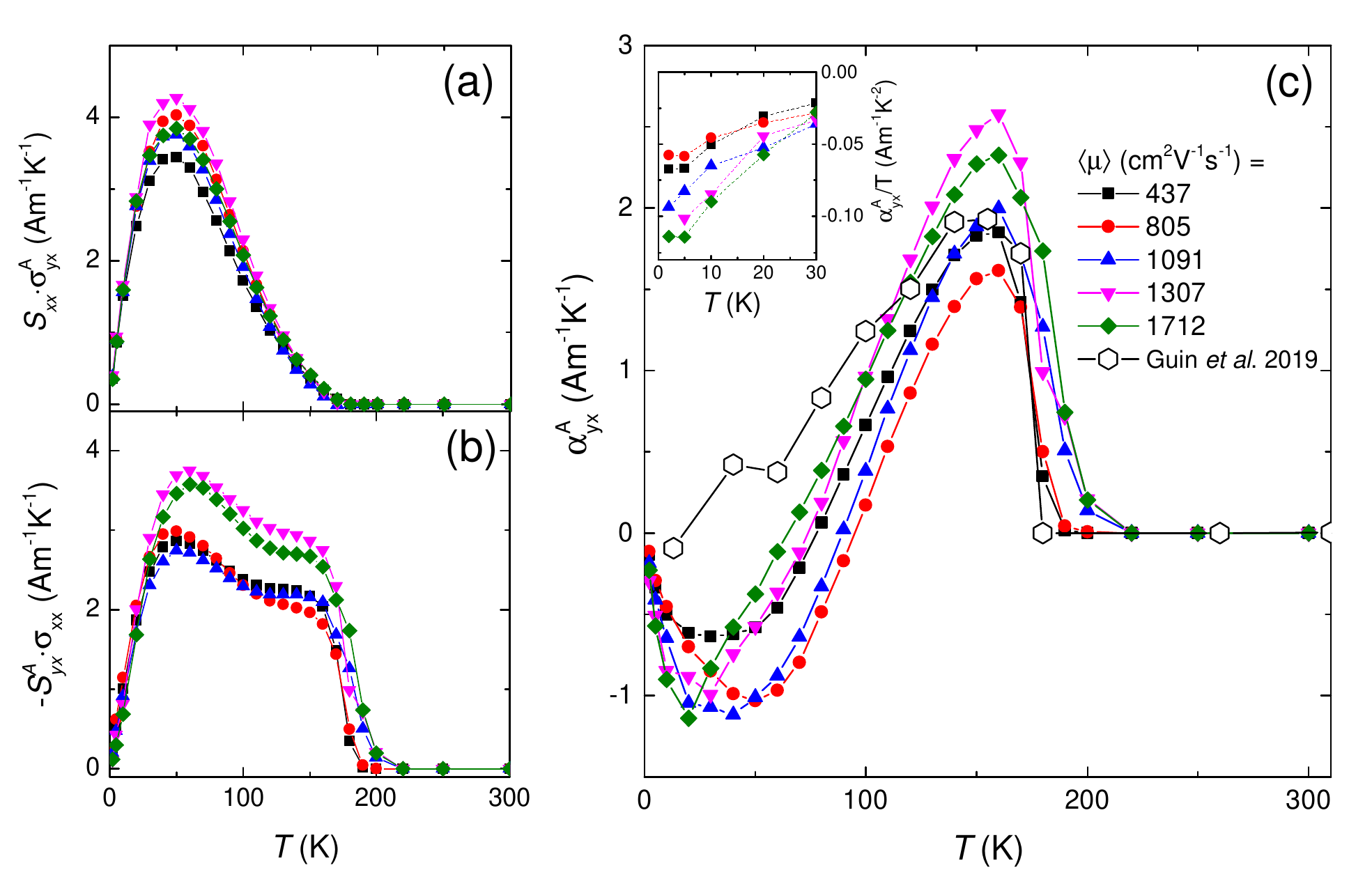}
\caption{\textbf{Anomalous off-diagonal component of the thermoelectric tensor $\alpha{_{yx}\rm ^A}$-} (a).(b). Temperature dependence of two products $S_{xx}\sigma{_{yx}\rm ^A}$ and $-S{_{yx}\rm ^A}\sigma_{xx}$.  The total $\alpha{_{yx}\rm ^A}$ is the difference between these two large signals. (c). The anomalous thermo-electricity coefficient  $\alpha{_{yx}\rm ^A}$ in the five samples.  Also shown is the data reported by Guin \textit{et al.}\cite{Guin2019}, Note the absence of the negative slope resolved here in their data. The inset shows the  $\alpha{_{yx}\rm ^A}/T$ at low temperatures to show the magnitude of the asymptotic negative slope in different samples.}
\label{alpha}
\end{figure*}

Fig. \ref{4coeff.} presents the temperature dependence of these four coefficients in the five samples. One can see how disorder affects the four different coefficients in different ways. As expected, low-temperature conductivity, $\sigma_{xx}$ is damped by decreasing mean-free-path (panel a). On the other hand, the low-temperature Seebeck coefficient, $S_{xx}$, does not very significantly between samples with different mobilities (panel b),  confirming that the chemical potential and doping level have not changed in any significant way. As we saw above, disorder affects mildly $\sigma{_{xy}\rm ^A}$ (panel c), but drastically $S{_{xy}\rm ^A}$ (panel d).

The two terms on the right side of Eq. \ref{10} are shown in Fig.\ref{alpha}a,b . As seen in the figure, the two terms are comparably large and have a maximum of the order of 3-4 Am$^{-1}$ K$^{-1}$. However they have opposite signs and as a result, $\alpha{_{xy}\rm ^A}$ is smaller and becomes negative below 80 K in all samples (See Fig. \ref{alpha}c). Our result is different from the two previous reports \onlinecite{Yuke2018,Guin2019}.

Ref.\onlinecite{Yuke2018} instead of subtracting the two components added them up. As a consequence, the peak amplitude of $\alpha{_{xy}\rm ^A}$ is much larger than what is found here or in ref. \onlinecite{Guin2019}. On the other hand, the data reported in ref. \onlinecite{Guin2019} matches our data over a temperature range but not below 100 K. In particular, Guin and co-workers did not resolve the sign change visible in five different samples studied in this work. We can only speculate that the sign change was missed because of the scarcity of their data points below 60 K (See Fig. \ref{alpha}c).

As one can see in the inset of Fig. \ref{alpha}c, the asymptotic zero-temperature slope of the anomalous off-diagonal thermoelectric conductivity $\alpha{_{xy}\rm ^A}/T$ in our samples is small and in the range of 0.05-0.10 Am$^{-1}$ K$^{-2}$. The dispersion among samples reflects the accumulated uncertainties in four distinct measurements and possibly some alight yet finite change in chemical potential. What we can firmly say is that the low-temperature $\alpha{_{xy}\rm ^A}$ is negative (and not positive).

\begin{figure*}
\includegraphics[width=18cm]{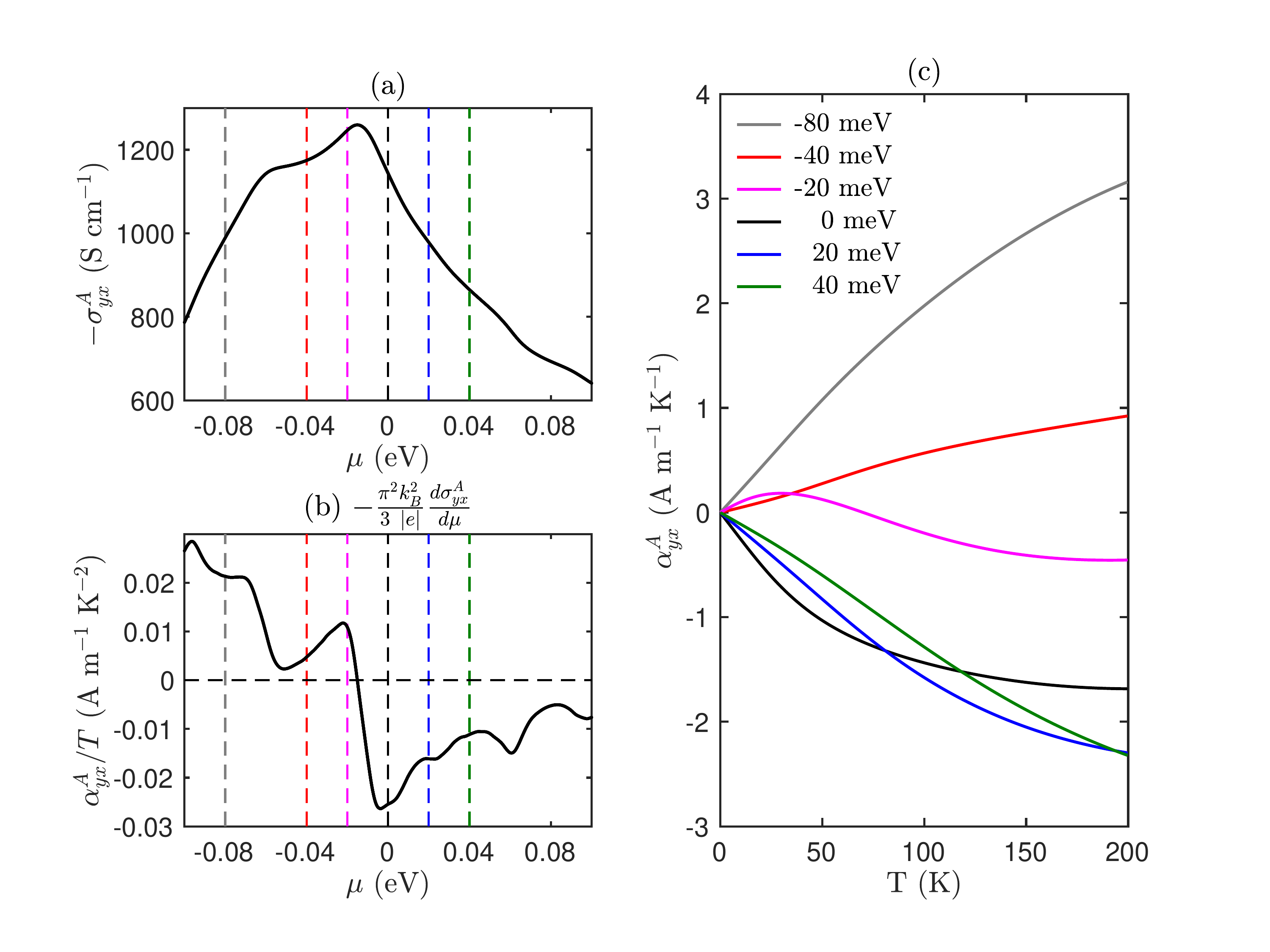}
\caption{\textbf{Theoretical $\sigma{_{yx}\rm ^A}$ and $\alpha{_{yx}\rm ^A}$}. (a) Anomalous Hall conductivity at 0 K as varying the chemical potential $\mu$. The charge neutral point is set to zero. (b) The slope of $\alpha_{yx}^A$ to $T$ in the Mott relation (Eq.~\ref{theory-Mottrelation}).
(c) Temperature dependence of $\alpha_{yx}^{A}$ at different chemical potentials.
}
\label{Fig:theory-Alpha}
\end{figure*}

In theory, we calculated the anomalous off-diagonal thermoelectric coefficient using \cite{Xiao2006}:
\begin{equation}
\alpha^A_{xy}(T,\mu)=-\frac{1}{e}\int d\epsilon \frac{\partial f(\epsilon-\mu,T)}{\partial \epsilon}\frac{\epsilon-\mu}{T} \sigma^A_{xy}(\epsilon),
\label{theory-alpha}
\end{equation}

Here $ \sigma^A_{xy}(\epsilon)$ is the chemical potential dependent AHC at the zero temperature (see Eq.~\ref{4}), $f(\epsilon-\mu, T)$ is the Fermi-Dirac distribution function and $\mu$ is the chemical potential. Near the zero temperature,  this equation gives the Mott relation:
\begin{equation}
\frac{\alpha_{xy}^A}{T} |_{T\rightarrow 0} = -\frac{\pi^2k_B^2}{3|e|} \frac{d\sigma^A_{xy}}{d\mu},
\label{theory-Mottrelation}
\end{equation}
We have calculated $\tilde\sigma^A_{yx}$ based on the Berry curvature extracted from the $ab~initio$ calculations and further obtained $\alpha^A_{yx}$ by Eq. ~\ref{theory-alpha}, as shown in Fig.~\ref{Fig:theory-Alpha}.
At the charge neutral point, $\alpha{_{yx}\rm ^A} $ exhibits nearly a peak slope to $T$ near the zero temperature. This large slope is negative and is induced by the large gradient of $-\sigma^A_{yx}$ (see Fig. \ref{Fig:theory-Alpha}a-b). Here $\alpha{_{yx}\rm ^A}/T$ is about --0.03 Am$^{-1}$ K$^{-2}$, slightly smaller than the experimental value. Actually the negative low-temperature slope exists in a wide chemical potential window ($\mu>-15$ meV).
This may explain why all five samples exhibit the negative slope, if there might be unnoticed small variation in the chemical potential.
We can get more insights from Eq.~\ref{8}.
Different from $\tilde\sigma^A_{xy}$, which is determined by the Berry curvature of all bands below the Fermi surface (see Eq.~\ref{4}), $\alpha_{yx}^A$ is a Fermi surface property and determined by the Berry curvature $\Omega_{yx}$ at the Fermi energy (Eq.~\ref{8}). As indicated in Fig.~\ref{fig:Fermisurface}, $\Omega_{yx}$ exhibits dominantly negative value near the charge neutral point, which is caused by anti-crossing bands (E1, E2 and H2 in the Fermi surface of Fig.~\ref{fig:Fermisurface}). Then large negative $\Omega_{yx}$ induces large negative $\alpha{_{yx}\rm ^A} /T $ at low temperature.

It is worth noting that $ab~initio$ calculations can only reveal the low-temperature physics. Present calculations do not include the effects such as the spin fluctuation and phonons at higher temperature, which can alter the the band structure and ANE. Thus,
it is not surprising that the calculated $\alpha{_{yx}\rm ^A} (T) $ does not show a sharp upturn to the positive side from about 40 K to the magnetic transition temperature as seen in Fig.~\ref{alpha}.

We note that Ref.~\onlinecite{Guin2019} invoked a shift of $-80$ meV in the chemical potential in order to explain a presumably positive $\alpha{_{xy}\rm ^A}$ at higher temperature. We note that given the small size of the Fermi surface pockets and their Fermi energies, such a shift in the chemical potential would imply a large departure from electron-hole compensation (see our supplementary information for the chemical potential dependent carrier densities), which would contradict their Hall conductivity data, which is similar to ours. We conclude that the agreement between our theory and experiment is such that there is no need to invoke a huge uncontrolled and unidentified doping to explain an  $\alpha{_{xy}\rm ^A}$ of opposite sign.

\section{Concluding remarks}
In summary, we performed an extensive set of electric and thermoelectric transport measurements on Co$_3$Sn$_2$S$_2$ with different impurity contents. We found that increasing the mean-free-path enhances the ordinary and reduces the anomalous components of the Nernst coefficients. Both theory and experiment find that this solid has a complex Fermi surface with multiple electron and hole pockets. However, the scaling of the anomalous Nernst coefficient with the inverse of the carrier mobility,  compatible with its intrinsic Berry-curvature origin, derived in a simple one-band approach persists. Our detailed analysis of the anomalous transverse thermoelectric conductivity, $\alpha{_{xy}\rm ^A}$, and its two components finds that it is negative at low temperature. Both its sign and amplitude match what theory expects without invoking uncontrolled doping.

\section{Acknowledgements}
This work was supported by the National Science Foundation of China (Grant No. 11574097 and No. 51861135104), the National Key Research and Development Program of China (Grant No.2016YFA0401704) and the Fundamental Research Funds for the Central Universities (Grant No. 2019kfyXMBZ071). Z. Z. was supported by the 1000 Youth Talents Plan. K. B. was supported by China High-end foreign expert program and by the Agence Nationale de la Recherche (ANR-18-CE92-0020-01).
B.Y. acknowledges the financial support by
the Willner Family Leadership Institute for the Weizmann Institute of Science,
the Benoziyo Endowment Fund for the Advancement of Science,
Ruth and Herman Albert Scholars Program for New Scientists, and the European Research Council (ERC) under the European Union¡¯s Horizon 2020 research and innovation programme (grant agreement No. 815869). H.C.L. was supported by the National Key R\&D Program of China (Grants No. 2016YFA0300504), the National Natural Science Foundation of China (No. 11574394, 11774423, 11822412).

\noindent
* \verb|binghai.yan@weizmann.ac.il|\\
* \verb|zengwei.zhu@hust.edu.cn|\\
* \verb|kamran.behnia@espci.fr|\\

\clearpage
\renewcommand{\thesection}{S\arabic{section}}
\renewcommand{\thetable}{S\arabic{table}}
\renewcommand{\thefigure}{S\arabic{figure}}
\renewcommand{\theequation}{S\arabic{equation}}

\setcounter{section}{0}
\setcounter{figure}{0}
\setcounter{table}{0}
\setcounter{equation}{0}


%
{\large\bf Supplemental Materials for ``Intrinsic anomalous Nernst effect amplified by disorder in a half-metallic semimetal"}
\section{Crystal growth and sample preparation}
High-quality single crystals (labelled with \#No.) of Co$_3$Sn$_2$S$_2$ were grown by chemical vapor transport (CVT) using iodine as the transport agent. Stoichiometric amounts of cobalt powder (99.9\%), tin powder (99.99\%), and sulfur pieces (99.99\%) in quartz ampule at 900$^\circ$C and 800$^\circ$C in a two-zone tube furnace for weeks. We also used a sample grown by tin-flux method \cite{LiuEnke2018SM,Lei2018SM} with lower residual resistivity ratio (RRR=$\rho(300$ K)/$\rho(2$ K)=4) labelled as B5. The properties of the samples studied including those shown in the main text is listed in table \ref{SM:Table:samples}.

\begin{table}[!hbp]
\begin{tabular}{c|c|c|c|c}
\hline
 Sample & RRR & growth method & A (n$\Omega$cmK$^{-2}$) & dimension (mm$^3$) \\
 \hline
B4 &4&flux&2.9&1.9$\times$1.1$\times$0.11\\
B5 &4&flux&3.2&1.5$\times$1.1$\times$0.12\\
\hline
\#6 & 16 &CVT&4.4&1.6$\times$1.4$\times$0.02\\
\#7 & 22 &CVT&4.5&1.2$\times$1.1$\times$0.007\\
\#12 & 8 &CVT&4.9&1.2$\times$0.9$\times$0.04\\
\#18 & 19 &CVT&4.5&1.9$\times$1.5$\times$0.2\\
\#21 & 22 &CVT&4.3&1.1$\times$0.7$\times$0.03\\
\#29 & 15 &CVT&4.4&1.3$\times$0.8$\times$0.02\\
\#30 & 10 &CVT&4.5&2.3$\times$1.4$\times$0.04\\
\#31 & 21 &CVT&4.3&1$\times$0.8$\times$0.003\\
\hline

\end{tabular}

\caption{Properties of the samples used in the resistivity measurements. CVT stands for chemical vapor transport growth method. Flux refers to the tin-flux growth method. $A$  is the prefactor of T-square resistivity in $\rho(T)=\rho_0+AT^2$.}
\label{SM:Table:samples}
\end{table}

\section{Hall and Nernst measurements}
Magnetoresistance and Hall resistivity were measured by the standard four-probe method using a current source with a DC nanovoltmeter in a Physical Property Measurement System (PPMS, Quantum Design). The Nernst measurements were performed using a chip-resistance heater and two pairs of thermocouples for the differential temperature measurement in PPMS under high-vacuum environment. The ordinary Nernst effect (ONE) $\nu/T$ is deduced from linearly fitting the field dependence of Nernst signal beyond the anomalous components identified by a jump in the Nernst signal. The anomalous Nernst (ANE) $S^A_{yx}$ is derived by the interception at $B=0$ from the linear part.

\begin{figure}
\includegraphics[width=9cm]{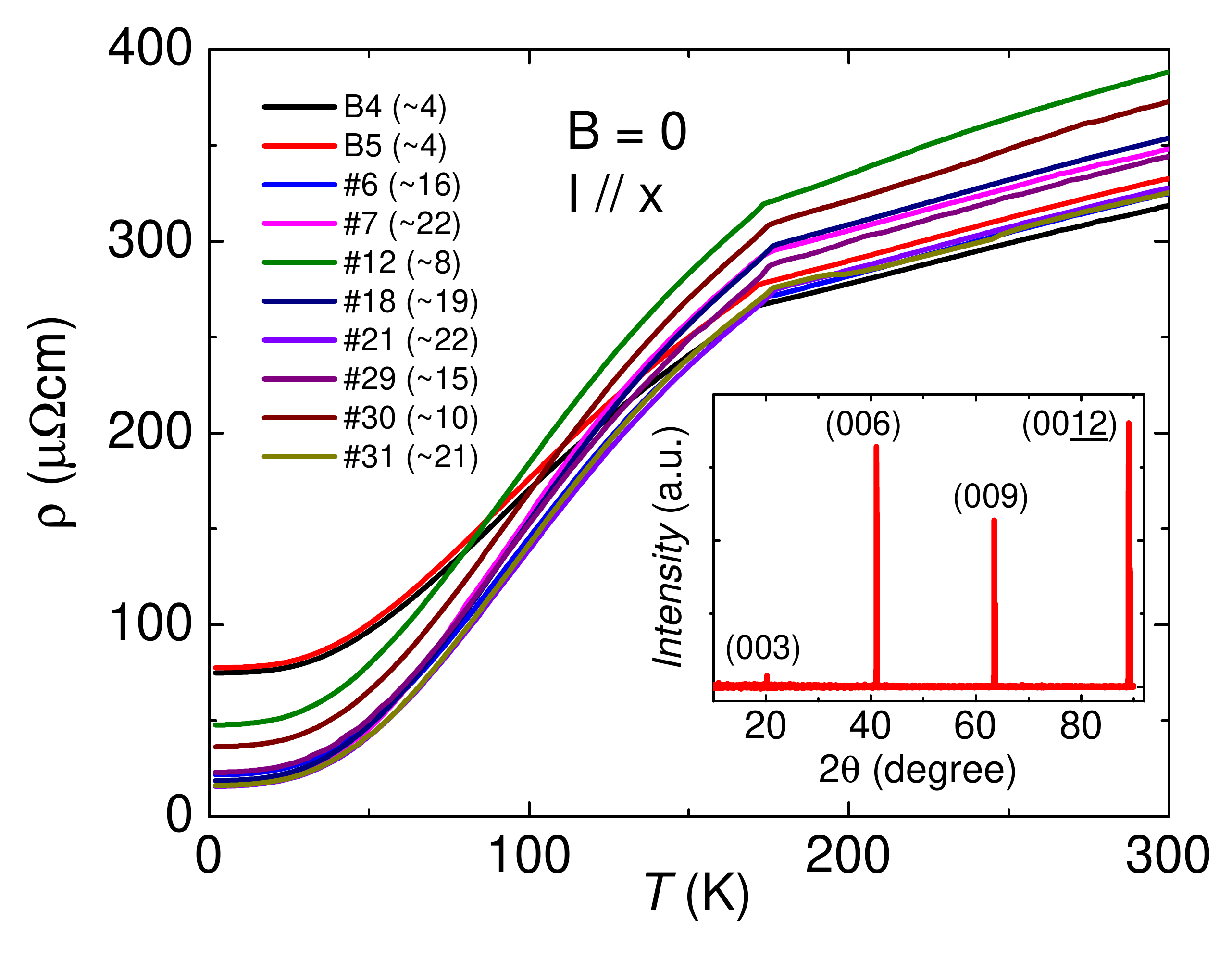}
\caption{Temperature dependence of resistivity for Co$_3$Sn$_2$S$_2$ samples at zero field. The inset shows sharp X-ray diffraction peaks, which can be
indexed by the indices of $(00l)$ lattice planes.}
\label{Fig: SMRT}
\end{figure}

The samples discussed in the main text were also examined by the energy dispersive spectroscopy (EDS) to determine their compositions. The results are listed in the table \ref{SM:Table:EDX}, which indicate almost same chemical characterization in these samples in the experimental margin of the EDS. Excess tin was observed on the surface of a the tin-flux as-grown sample Fig. \ref{Fig: SMEDX}b. However, its EDS result  after polishing is comparable with others. We notice that a similar Sn excess of  2.07 in all samples. This slight departure from perfect stoichiometry does not seem to correlate with the slight imbalance between electrons and holes seen in the Hall data.

\begin{figure}
\includegraphics[width=9cm]{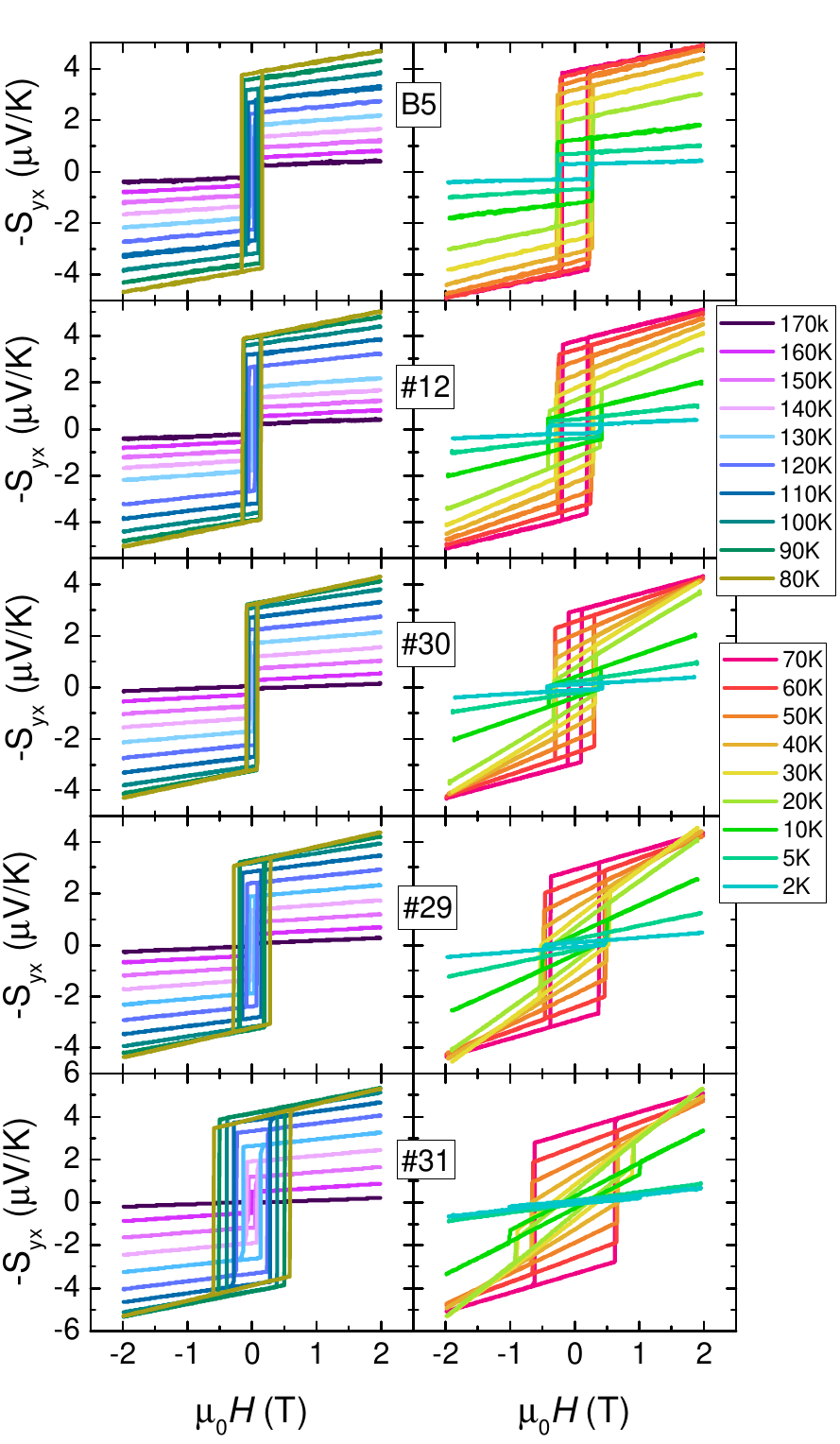}
\caption{Temperature-dependent Nernst response of Co$_3$Sn$_2$S$_2$ samples for a magnetic field applied along the z-axis.}
\label{Fig: NERNST}
\end{figure}

\begin{figure}
\includegraphics[width=9cm]{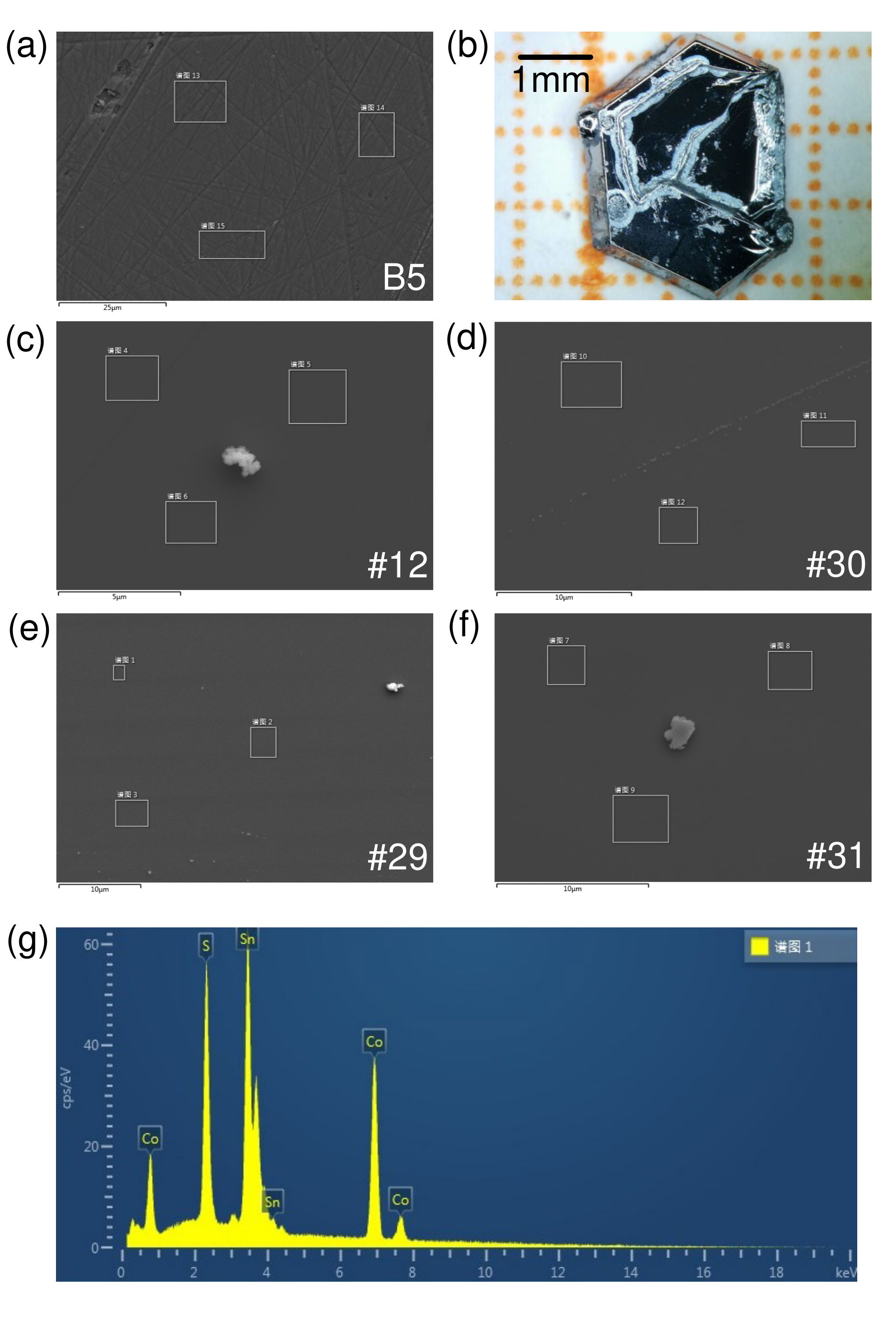}
\caption{ (a)(c)(d)(e)(f). The scanning electron microscope (SEM) images for the samples discussed in the main text. (b). Image of a tin-flux as-growth sample. Excess tin can be seen clearly on the surface of the sample. (g). The energy dispersive spectroscopy (EDS) spectrum of one of the Co$_3$Sn$_2$S$_2$ samples. }.
\label{Fig: SMEDX}
\end{figure}

\begin{table}[!hbp]
\begin{tabular}{|c|c|c|c|c|}

\hline
 Sample & Cobalt & Tin  & Sulfur \\
\hline
B5 & 3.01 & 2.07 & 2.00 \\
\hline
\#12 & 3.01 & 2.07 & 1.99 \\
\hline
\#30 & 3.02 & 2.07 & 1.98 \\
\hline
\#29 & 3.00 & 2.07 & 2.00 \\
\hline
\#31 & 2.99 & 2.07 & 2.00 \\
\hline

\end{tabular}
\caption{Elemental ratios in Co$_3$Sn$_2$S$_2$ samples from EDS, showing a consistent chemical composition result. }
\label{SM:Table:EDX}
\end{table}

\section{Carrier density and mobility }

\begin{table}[!hbp]
\begin{tabular}{|c|c|c|c|c|c|c|c|c|}
\hline
 Sample & RRR & Grown &n$_h$ & n$_e$ & $\mu_h$ & $\mu_e$& $\langle\mu\rangle$\\
\hline
B5 &4&flux&8.74&8.78&399&397&437\\
\hline
\#12 & 8 &CVT&8.74&8.72&802&801&805\\
\hline
\#30 & 10 &CVT&8.73&8.68&1003&1001&1091\\
\hline
\#29 & 15 &CVT&8.91&8.71&1377&1386&1307\\
\hline
\#31 & 21 &CVT&8.77&8.42&1768&1796&1712\\
\hline
\end{tabular}
\caption{ Electron (hole) arrier density [ n$_e$ (n$_h$)] (in units of 10$^{19}$cm$^{-3}$) together with their mobilities [$\mu_e$ ($\mu_h$)] (in units of cm$^2$V$^{-1}$s$^{-1}$) in different samples. They were extracted using a two-band model of the ordinary Hall conductivity. The average mobility, $\langle\mu\rangle$, is extracted from fitting magnetoresisitance by $(\rho(B)-\rho(0))/\rho(0)\propto (\langle \mu \rangle B)^2$.}
\label{Table:mobility}
\end{table}

The Hall resistivity contains two components: $\rho_{H} = R_{0}B + 4\pi R_{s}M$ where $R_{0}B$ is the normal part and  $4\pi R_{s}M$ is the anomalous one. Therefore, the ordinary component is $\rho^N_{yx}(B)=\rho_{yx}- 4\pi R_{s}M$ and the ordinary Hall conductivity $\sigma^N_{yx}=-\rho^N_{yx}/(\rho^N_{yx}{}^2+\rho_{xx}^2)$. Before fitting the ordinary Hall conductivity, we subtracted the anomalous part from the total Hall resistivity.

We employed two methods to extract the mobility in each sample. First, a fit to the field dependence of the ordinary Hall conductivity using a two-band model:
\begin{equation}
 \sigma_{yx} = [\frac{n_{h}\mu_h^2}{1 + \mu^2_{h}B^2}-\frac{n_{e}\mu_e^2}{1 + \mu^2_{e}B^2}]eB
\end{equation}

Here; $\sigma_{yx}$ is the Hall conductivity and B is the magnetic field. $n_{h}$($n_{e}$) is the carrier density of holes (electrons), $\mu_{h}$($\mu_{e}$) is the carrier mobility of holes (electrons).

The fitting results for carrier density and mobility are listed in the table\ref{Table:mobility}.
As seen in the table, the samples are close to perfect compensation.

Low-field magnetoresistance is another way to extract mobility, given that $(\rho(B)-\rho(0))/\rho(0)\propto (\langle \mu \rangle B)^2$. The results are also listed in the table \ref{Table:mobility}. Mobilities extracted using these two methods  are close to each other.

\begin{figure}
\includegraphics[width=9cm]{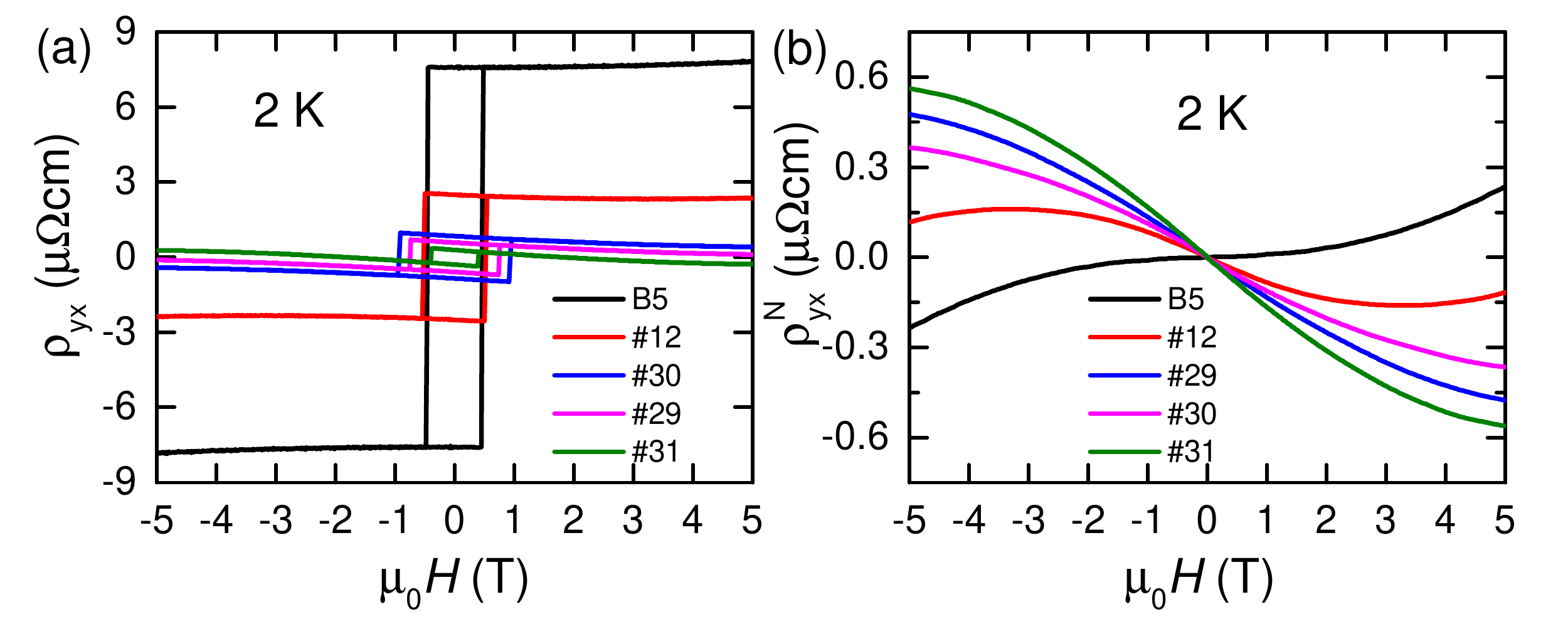}
\caption{ (a)Field dependence of Hall resistivity and (b)the ordinary components of Hall resistivity at 2 K on five samples. }.
\label{Fig: OHE}
\end{figure}

\section{T-square resistivity}

\begin{figure}
\includegraphics[width=9cm]{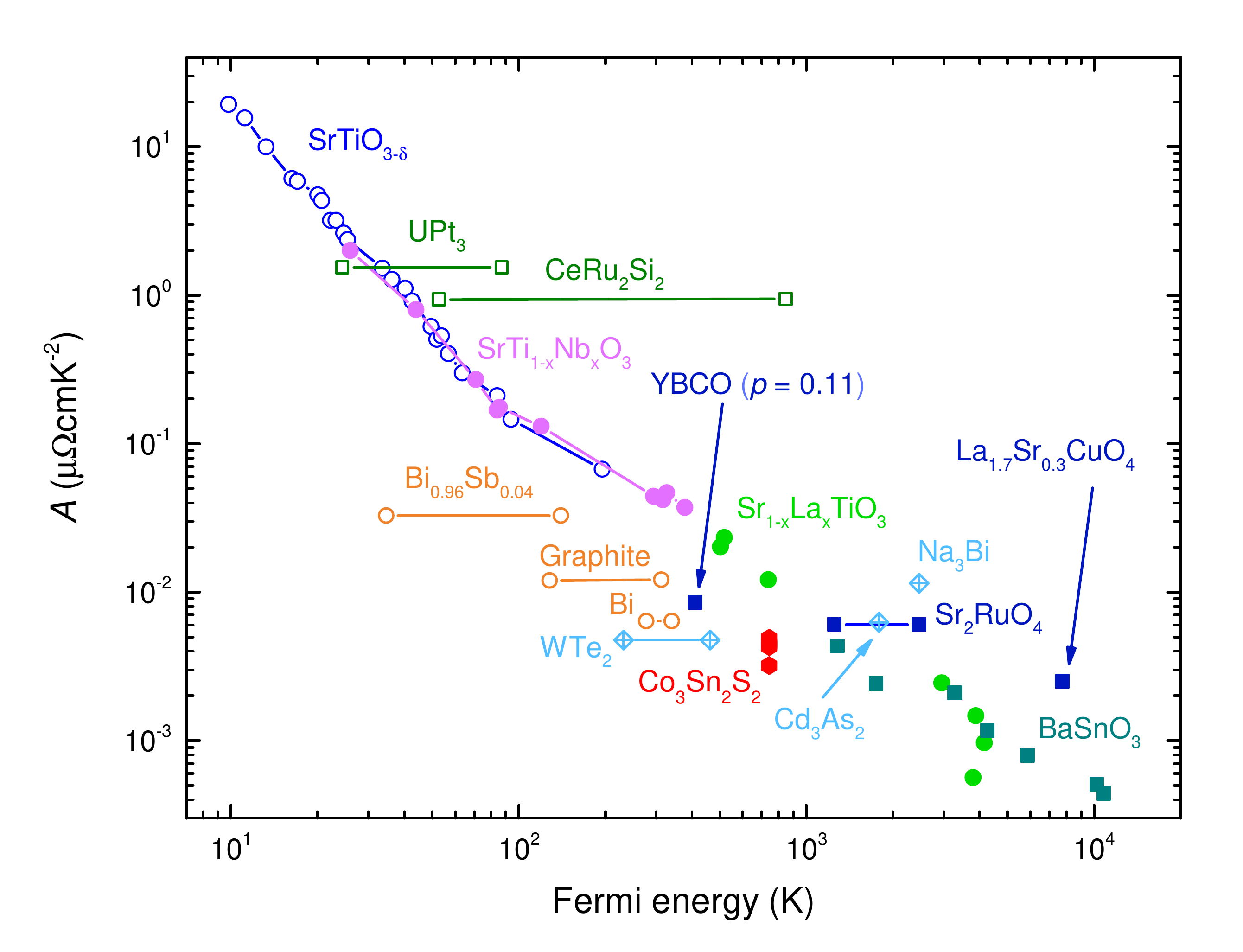}
\caption{ The prefactor $A$ of $T^2$ resistivity $vs.$ Fermi temperature for a number of dilute metals\cite{Clement2019SM} with our Co$_3$Sn$_2$S$_2$ data included. }
\label{Fig: A_EF}
\end{figure}

The resistivity of a Fermi liquid is expected to follow a quadratic temperature dependence: ($\rho(T)=\rho_0+\rm{A}T^2$). The prefactor $A$ is linked to the Fermi energy. The $T^2$ prefactor $A$ in  Co$_3$Sn$_2$S$_2$ samples are listed in table \ref{SM:Table:samples}. As seen in Fig. \ref{Fig: A_EF}, which shows the data for numerous dilute metals and find the consistence of our data with these metals, pointing to a universal law for the behavior between A and E$_F$. The magnitude of $A$ and Fermi energy in Co$_3$Sn$_2$S$_2$  correlate as observed in other dilute metals \cite{Clement2019SM}.


\section{Derivation of $\alpha_{xy}$ using the Mott formula}

According to the Mott relation, $\alpha_{ij}$ is the energy derivative of $\sigma_{ij}$ :

\begin{equation}\label{S1}
\alpha_{ij}= -\frac{\pi^2}{3|e|}k_B^2T \frac{\partial \sigma_{ij}(\epsilon)}{\partial\epsilon }|_{\epsilon=\epsilon_F}
\end{equation}

As stated in the main text, the low-field expression for two-dimensional Hall conductivity is :
\begin{equation}\label{S2}
\sigma^{LF}_{xy} \approx \frac{e^2}{h} \frac{\ell^{2}}{\ell_B^{2}}
\end{equation}

When the mean-free-path is energy-independent, the energy derivative of $\sigma^{LF}_{xy}$ is zero. A more realistic assumption is some  $\ell =\ell_0  (1+ \epsilon^{\alpha}/\epsilon_F$), with $\alpha \approx 1$. Therefore:
\begin{equation}\label{S3}
 \frac{\partial \ell(\epsilon)}{\partial\epsilon }|_{\epsilon=\epsilon_F} \approx \frac{\ell}{2\epsilon_F}
\end{equation}

This yields:
\begin{equation}\label{S4}
\alpha^{LF}_{xy}= \frac{\pi^2}{3} \frac{ek_B}{h}  \frac{2\ell^{2}}{\ell_B^{2}}\frac{k_BT}{\epsilon_F}
\end{equation}
By replacing $\frac{ k_BT}{\epsilon_F}$ with $\frac{\lambda_F^2}{\Lambda^2}$, one deduces Eq. 5 of the main text.

In the high-field regime, the expression for two-dimensional Hall conductivity is :
\begin{equation}\label{S5}
\sigma^{HF}_{xy} (\epsilon_k) \approx \frac{e^2}{h} \ell_B^{2} k^{2}
\end{equation}

 With a parabolic dispersion ($\epsilon=\frac{\hbar^2k^2}{2m^*}$), one has:
\begin{equation}\label{S3}
\frac{\partial k}{\partial\epsilon }|_{\epsilon=\epsilon_F}=
\frac{m^*}{\hbar^{2}k_F }
\end{equation}

Using Eq. S1 and $\Lambda^2=\frac{h^2}{2 \pi m^*k_BT}$, one finds:

\begin{equation}\label{S4}
\alpha^{HF}_{xy} \approx  \frac{ek_B}{h}  \frac{\ell_B^{2}}{\Lambda^{2}}
\end{equation}

Note that we are neglecting numerical factors of the order of unity. This is the expression given in the Eq. 7 of the main text.

\section{DFT calculation}
Our calculations were performed using the density-functional theory (DFT) in the framework of the generalized gradient approximation\cite{PBE1996}) with the full-potential local-orbital minimum-basis code (FPLO)\cite{Koepernik1999}.
Spin-orbit coupling was included in all calculations.
We obtained atomic like Wannier functions from DFT Bloch wave functions. Then we built the Wannier-based tight-binding Hamiltonian and calculated the Berry curvature and anomalous Hall conductivity in the clean limit. We employed a k-point grid of $300\times300\times300$ for the anomalous Hall conductivity.)

\begin{figure}
\includegraphics[width=9cm]{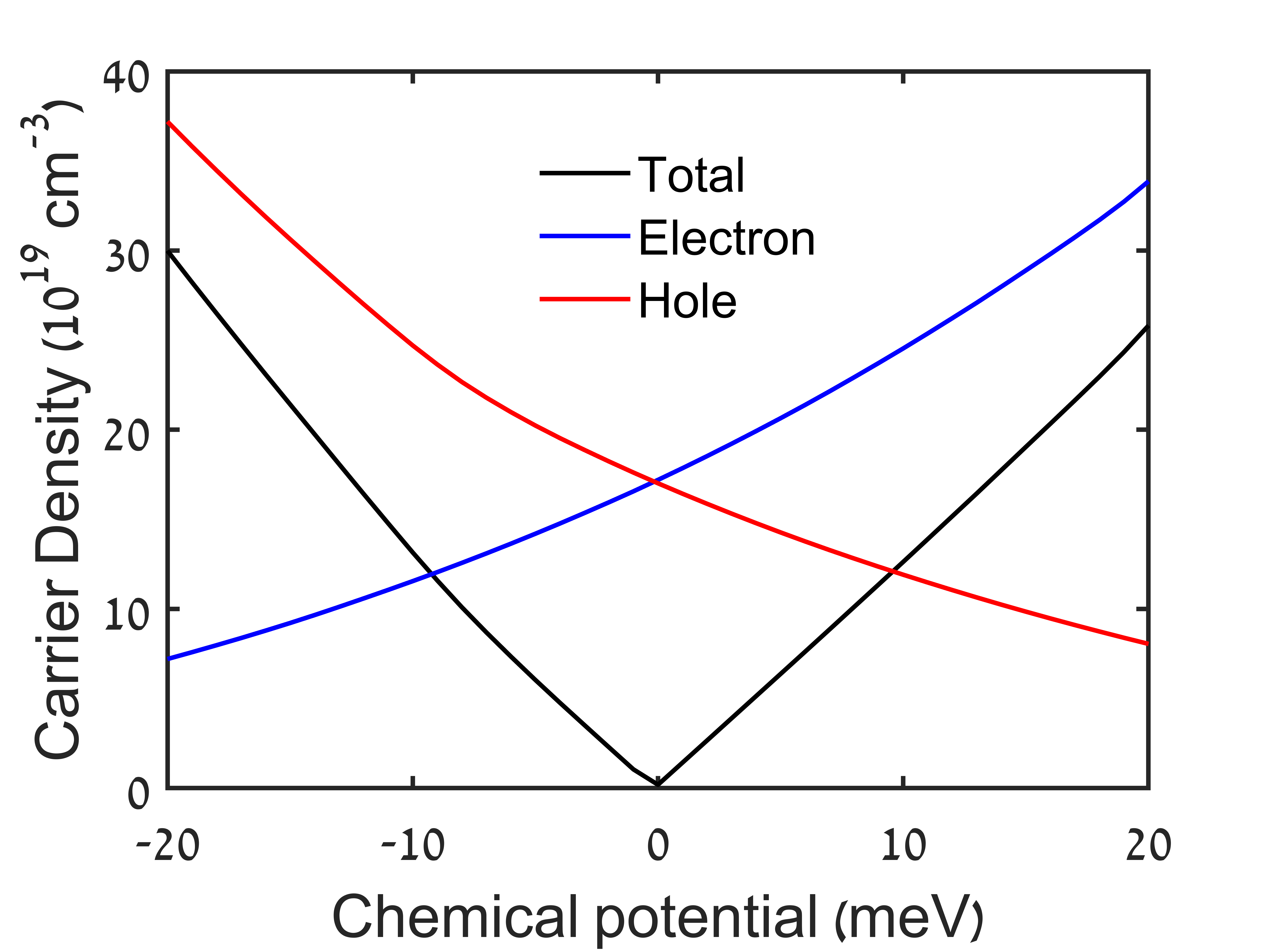}
\caption{ The electron and hole carrier densities as a function of the chemical potential.
The electron/hole densities are calculated by integrating the volume of all related Fermi pockets. The charge neutral point is set the exact electron-hole compensation point. We note that the charge neutral point estimated from ordinary DFT density of states is usually less accurate with several meV error bar because of the sparse k-point grid.}
\label{Fig: Carrier density from theory}
\end{figure}

\begin{figure}
\includegraphics[width=9cm]{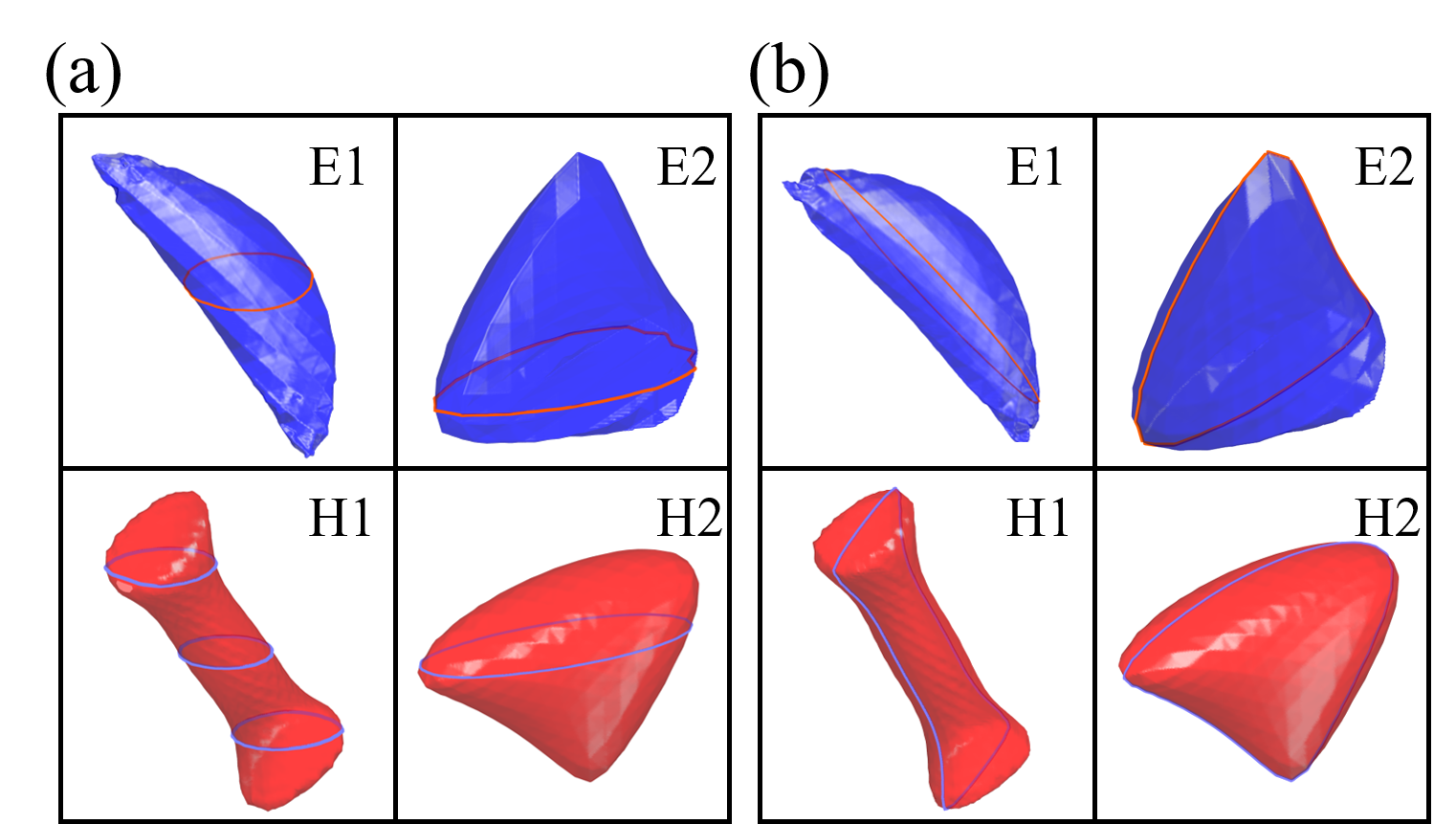}
\caption{ The shape of each Fermi pockets and related extreme orbits for (a) B//\textit{z} and (b) B//\textit{xy}.  From the cross-section area of Fermi pocket, we evaluate SdH frequencies and effective masses in table. \ref{Table:FS}}
\label{Fig: ExOrbit}
\end{figure}

\end{document}